\documentclass[prd,superscriptaddress,amsfonts,amssymb,amsmath,showpacs,onecolumn]{revtex4-2}
\usepackage{bm}
\usepackage{amsfonts}
\usepackage{latexsym}
\usepackage[latin1]{inputenc}
\usepackage{graphicx}
\usepackage{amsmath}
\usepackage{palatino}
\usepackage{mathpazo}
\usepackage{textcomp}
\usepackage{float}
\usepackage{booktabs}
\usepackage{dcolumn}
\usepackage{multirow}
\usepackage{ragged2e}
\usepackage{hyperref}
\hypersetup{colorlinks,citecolor=blue}
\usepackage{amsmath}
\usepackage{xcolor}
\usepackage{orcidlink}
\usepackage[caption=false]{subfig}
\usepackage{commath}
\def\jnl@style{\it}
\def\aaref@jnl#1{{\jnl@style#1}}

\def\aaref@jnl#1{{\jnl@style#1}}

\def\aj{\aaref@jnl{AJ}}                   
\def\apj{\aaref@jnl{ApJ}}                 
\def\apjl{\aaref@jnl{ApJ}}                
\def\apjs{\aaref@jnl{ApJS}}               
\def\apss{\aaref@jnl{Ap\&SS}}             
\def\aap{\aaref@jnl{A\&A}}                
\def\aapr{\aaref@jnl{A\&A~Rev.}}          
\def\aaps{\aaref@jnl{A\&AS}}              
\def\mnras{\aaref@jnl{Mon.~Not.~Roy.~Astron.~Soc.}}             
\def\prd{\aaref@jnl{Phys.~Rev.~D}}        
\def\prc{\aaref@jnl{Phys.~Rev.~C}}  
\def\prl{\aaref@jnl{Phys.~Rev.~Lett.}}    
\def\qjras{\aaref@jnl{QJRAS}}             
\def\skytel{\aaref@jnl{S\&T}}             
\def\ssr{\aaref@jnl{Space~Sci.~Rev.}}     
\def\zap{\aaref@jnl{ZAp}}                 
\def\nat{\aaref@jnl{Nature}}              
\def\aplett{\aaref@jnl{Astrophys.~Lett.}} 
\def\apspr{\aaref@jnl{Astrophys.~Space~Phys.~Res.}} 
\def\physrep{\aaref@jnl{Phys.~Rep.}}      
\def\physscr{\aaref@jnl{Phys.~Scr}}       
\def\commat{\aaref@jnl{Comm.~Math.~Phys.}}              
\def\science{\aaref@jnl{Science}}               
\def\cqg{\aaref@jnl{Classical Quant.~Grav.}}            
\def\jpcs{\aaref@jnl{JPCS}}                                     
\def\ijmpd{\aaref@jnl{Int.~J.~Mod.~Phys.~D}}                    
\def\grg{\aaref@jnl{Gen.~Relat.~Gravit.}}               
\def\rpp{\aaref@jnl{Rep.~Prog.~Phys.}}          
\def\npa{\aaref@jnl{Nucl.~Phys.~A}}        
\def\lrr{\aaref@jnl{Living Rev.~Rel.}}                   
\def\jcap{\aaref@jnl{J.~Cosmology Astropart.~Phys.}}    
\def\rmp{\aaref@jnl{Rev.~Mod.~Phys.}}   
\def\epjc{\aaref@jnl{Eur.~Phys.~J.~C}}

\allowdisplaybreaks[1]

\begin{document}

\color{black}       

\title{Polytropic $f(Q)$ cosmology and its implications for the $H_0$ tension}

\author{Raja Solanki\orcidlink{0000-0001-8849-7688}}
\email{rajasolanki8268@gmail.com}
\affiliation{Department of Mathematics, School of Advanced Sciences,\\ VIT-AP University, Beside AP Secretariat, Amaravati, 522241, Andhra Pradesh, India.}


\date{\today}
\begin{abstract}
Understanding the late-time cosmic phenomenon of the universe commonly referred to as the dark energy problem, which is one of the prominent tension in the field of theoretical as well as observational cosmology. In this work, we attempt to analyze the nature of the missing fluid of the universe. In order to do so, we employ a polytropic equation of state consisting of free parameters rather assuming directly a particular form of the fluid. In addition, for the background geometry we consider a $f(Q)$ cosmology exhibiting power-law assumption, which is recently proposed and found to be attractive in the study of late-time cosmology. We find exact cosmological solution along with a rigorous data analysis, utilizing the Bayesian statistics approach and the emcee ensemble sampler, to find the parameter constraints and then we interpret the parameters of physical interests such as deceleration and statefinder parameter. Also, we present status of the $H_0$ tension predicted by our polytropic $f(Q)$ cosmological model.
\end{abstract}

\maketitle

\textbf{Keywords:} non-metricity, $f(Q)$ gravity, polytropic equation of state, and $H_0$ tension.  

\section{Introduction}\label{sec1}
\justifying
Determining the precise value of $H_0$ has been a significant challenge, as it was obtained by measuring large cosmic distances. At first, Hubble estimated a value of $H_0$ as $ 500 \: km/s/Mpc$ \cite{R1}, which was overestimated due to inaccuracies in distance calibrations. After the launch of the Hubble Space Telescope (HST), the range of $H_0$ value was approximately between $50-100 \: km/s/Mpc$. During the last decade, the Planck collaboration obtained the $H_0$ value as $67.4 \pm 0.5 \: km/s/Mpc $ (Planck18) by analyzing the cosmic microwave background radiation (CMB) data \cite{Planck2020}. 
Other measurements of the $H_0$ value that align with the Planck results i.e. favoring the lower values of $H_0$ are the ground-based CMB telescope, the Wilkinson Microwave Anisotropy Probe (WMAP) + Atacama Cosmology Telescope (ACT-DR4) \cite{R3} estimates a value of $H_0$ as $67.6 \pm 1.1 \: km/s/Mpc$ at $68$\% confidence level (CL) in $\Lambda$CDM, whereas the South Pole Telescope (SPT-3G) reports $H_0 = 68.8 \pm 1.5 \: km/s/Mpc $ at the same CL \cite{R4}. One of the remarkable issues in the era of modern cosmology is the $H_0$ tension (for instance, one can check reviews \cite{PL6,PL7}), which is the disagreement in $H_0$ values extracted from the local measurements via the SH0ES collaboration, for instance, R19 \cite{R6} reports $H_0 = 74.03 \pm 1.42 \: km/s/Mpc $ at a $68$\% CL, or R20 \cite{R7} when assuming Gaia EDR3 parallax measurements, providing $H_0 = 73.2 \pm 1.3 \: km/s/Mpc $ at the same CL, and that of the Planck \cite{Planck2020} estimate $H_0 = 67.27 \pm 0.60 \: km/s/Mpc$ at $68$\% CL, invoking the $\Lambda$CDM as the base cosmological model.

Gravity can be understood as a geometric phenomenon arising from the universal character of the equivalence principle. In standard General Relativity, this geometric description is expressed through the curvature of spacetime, which has become the conventional way to interpret gravitation. Nevertheless, curvature is not the only framework available. Two other formulations that are dynamically equivalent to General Relativity but are constructed in flat spacetime. In these approaches, gravity is attributed entirely either to spacetime torsion or to non-metricity. This reveals that the same fundamental theory can be represented in three different, apparently independent, geometric languages called the trinity of gravity \cite{fQfT}. In recent years, gravitational frameworks incorporating non-metricity have drawn considerable attention \cite{Nester,jimenez2,fQfT,Runkla:2018xrv,jimenez1}.
These theories employ the non-metricity scalar $Q$ within the Palatini formalism, where the metric and the affine connection are considered as independent entities. When the Riemann tensor encoding the description of spacetime curvature and the torsion tensor of the connection are treated as non-contributing, the connection can be described with the help of four scalar fields \cite{DBLXT,BeltranJimenez,Adak,Tomonari}.
A particular choice of these scalar fields allows this connection to vanish, and the corresponding gauge is referred to as the coincident gauge \cite{FDA}, under which the metric tensor becomes the only dynamical variable. In this gauge, a linear action in the non-metricity $Q$ dynamically corresponds to general relativity (GR), a formulation referred to as the Symmetric Teleparallel Equivalent to GR (STEGR) \cite{coincident,Quiros}. While STEGR reproduces the dynamics of GR, its reformulation in terms of non-metricity opens a new route for constructing gravitational extensions. By replacing the linear dependence on $Q$ with a general function $f(Q)$, one can obtain a modified theory of gravity that preserves the second-order characteristic of the field equations while introducing richer phenomenology \cite{jimenez1,Lazkoz2019,Khyllep2023}. Such $f(Q)$ models have been actively explored in recent cosmological studies as promising candidates to resolve the prominent mystery of the accelerating universe without adhering to dark energy in explicit form. Observational studies using SNe Ia, BAO, cosmic chronometers, CMB distance priors, and gravitational-wave data suggest that parametric forms of $f(Q)$ such as exponential, power-law, and logarithmic models for extensions of STEGR \cite{PL8,PL9,PL10,PL11,PL12,PL13,PL14,PL15,PL16,PL17,PL18} are consistent with cosmological observations.

On the other hand, from the theoretical perspective, different fluids such as barotropic fluid, chaplygin gas, viscous dark energy, different scalar field model such as quintessence, power-law potential, exponential potential, have been tested against the observation data. The precision in observational cosmology indicates that none of our model is capable of drawing complete picture of the universe without any tension. However, each of these proposed plethora of theories is well sound in the sense of describing the piece of evolution, such as inflation, the early universe, the post-inflationary era, nucleosynthesis, and the late-time cosmology. Moreover, it is required to the upcoming generations to enhance our statistical understanding and precision along with the theoretical investigation. The polytropic equation of state (EoS) is generally characterized the stiffness and non-linearity of the fluid. In relativistic contexts, this relation reduces to the well-known polytropic models applied to stellar structures, linking microscopic thermodynamics with large-scale fluid behavior \cite{Tooper1964}. The relativistic form of the polytrope, developed in detail by Tooper, remains a standard reference for describing compact astrophysical objects. In cosmology, this class of models has been extended and generalized to include mixed linear-polytropic forms and negative-index cases, enabling a unified description of cosmic evolution that transitions between dust-like, radiation-like, and accelerated expansion phases \cite{Chavanis2012}. Chavanis and others have shown that suitable choices of parameters can produce late-time cosmic acceleration or even avoid singularities altogether. Earlier studies have also explored polytropic gas as a dark energy candidate, uncovering behaviors such as transient phantom phases, thermodynamic consistency, viscous effects, and scalar field reconstructions \cite{Mukhopadhyay2005,Moradpour2015}. Recently, the coupling of polytropic matter to $f(Q)$ gravity was studied in astrophysical contexts, with Tolman-Oppenheimer-Volkoff-like equations solved for compact star configurations. These studies reveal that the presence of a polytropic EoS in $f(Q)$ gravity can shift mass-radius relations, alter maximum mass predictions, and modify anisotropy and complexity measures compared to general relativity \cite{LinZhai2021,deAraujo2024}. Such findings emphasize the relevance of combining the polytropic EoS with $f(Q)$ models for cosmological modeling in order to relieve the $H_0$ tension or providing an alternative mechanism to $\Lambda$CDM to recreate the expansion history. In the introductory text section \ref{sec1} of the article, we highlight key fundamentals on the modified symmetric teleparallel theory along with a thorough literature review on the polytropic equation of state and the $H_0$ tension. In section \ref{sec2} of the article we present mathematical foundations of the $f(Q)$ cosmology. Then we explore the polytropic equation of state in section \ref{sec3} and then we find the analytical model solution. Further in section \ref{sec4}, we discuss the data methodology require for the model parameter estimation. We then thoroughly discussed the obtain results in section \ref{sec5}. At the end, in the section \ref{sec6}, we close our article by highlighting concluding remarks on the investigation performed by us.

\section{$f(Q)$ Cosmology}\label{sec2}
\justifying
General relativity, in its traditional form, is constructed within the framework of Lorentzian geometry, relying on a connection, which is symmetric in nature and exhibits the metricity condition, known as the Levi-Civita connection. This particular connection inherently gives rise to curvature, while ensuring that torsion remains absent \cite{fQfT}. Alternatively, one may consider the Weitzenb$\ddot{o}$ck connection, which, although still compatible with the metric, allows for torsion. This approach leads to what is identified as the teleparallel formulation of general relativity (TEGR). Beyond these, the most generalized form of a metric-affine connection is described by the below expression:
\begin{equation}\label{2a}
\hat{\Gamma}^{\,\sigma}_{\,\,\,\alpha\beta}=\Gamma^{\,\sigma}_{\,\,\,\alpha\beta}+K^{\,\sigma}_{\,\,\,\alpha\beta}+L^{\,\sigma}_{\,\,\,\alpha\beta}
\end{equation}
In this expression, the first term corresponds to the Levi-Civita connection, given explicitly as\\ $\Gamma^{\,\sigma}_{\,\,\,\alpha\beta}=\frac{1}{2}g^{\sigma\lambda}\left(\partial_{\alpha}g_{\lambda\beta}+\partial_{\beta}g_{\lambda\alpha}-\partial_{\lambda}g_{\alpha\beta}\right)$, which is symmetric in the indices $\alpha$ and $\beta$, and adhering the metricity condition. The second term denotes the contorsion tensor, calculated as $K^{\,\sigma}_{\,\,\,\alpha\beta}=\frac{1}{2}T^{\,\sigma}_{\,\,\,\alpha\beta}+T^{\,\,\,\,\,\,\,\sigma}_{(\alpha\,\,\,\,\,\,\beta)}$, where $T^{\,\sigma}_{\,\,\,\alpha\beta}$ is the torsion tensor defined in terms of Weitzenb$\ddot{o}$ck connection as  $T^{\,\sigma}_{\,\,\,\alpha\beta} = W^{\,\sigma}_{\,\,\,\beta\alpha}- W^{\,\sigma}_{\,\,\,\alpha\beta}$, which is antisymmetric in the indices $\alpha$ and $\beta$, while the third term, known as the disformation tensor, exhibiting the form $L^{\,\sigma}_{\,\,\,\alpha\beta}=-\frac{1}{2}g^{\sigma\lambda}\left(Q_{\alpha\lambda\beta}+Q_{\beta\lambda\alpha}-Q_{\lambda\alpha\beta}\right)$.\\
The symmetric teleparallel formulation of general relativity (STEGR) was developed by employing a symmetric connection that exhibits vanishing torsion while violating metric compatibility. In this formulation, gravitational phenomenon are encoded in the resulting non-metricity tensor, which is given below,
\begin{equation}\label{2b}
Q_{\sigma\alpha\beta}=\nabla_{\sigma}g_{\alpha\beta},
\end{equation}
The associated trace components of the non-metricity tensor are presented as below,
\begin{equation}\label{2c}
Q_{\sigma}=Q_{\sigma\,\,\,\,\alpha}^{\,\,\,\,\alpha}\, ,\,\,\,\,\,\,\,\,\tilde{Q}_{\sigma}=Q^{\alpha}_{\,\,\,\,\sigma\alpha}\,.
\end{equation}
Moreover, the corresponding conjugate of the non-metricity tensor is acquired as,
\begin{equation}\label{2d}
4P_{\,\,\alpha\beta}^{\sigma}=-Q^{\sigma}_{\,\,\,\,\alpha\beta}+2Q^{\,\,\,\,\,\,\sigma}_{(\alpha\,\,\,\,\beta)}+Q^{\sigma}g_{\alpha\beta}-\tilde{Q}^{\sigma}g_{\alpha\beta}-\delta^{\sigma}_{(\alpha}\, Q\,_{\beta)},
\end{equation}
The non-metricity scalar is then obtained as,
\begin{equation}\label{2e}
Q=-Q_{\sigma\alpha\beta}P^{\sigma\alpha\beta}.
\end{equation}
The general action describing $f(Q)$ gravity, including two Lagrange multiplier terms, is formulated as \cite{jimenez2},
\begin{equation}\label{2f}
S=\int \left[\frac{1}{2}f(Q)+\lambda_{\alpha}^{\,\,\,\beta\mu\nu} R^{\alpha}_{\,\,\,\beta\mu\nu}+\tilde{\lambda}_{\alpha}^{\,\,\,\mu\nu} T^{\alpha}_{\,\,\,\mu\nu}+\mathcal{L}_m\right]\sqrt{-g}\,d^4x,
\end{equation}
Here, $f(Q)$ denotes a function of the non-metricity scalar, $g = \det(g_{\alpha\beta})$ represents the determinant of the metric tensor, $\mathcal{L}_m$ is identified as the matter Lagrangian density, and $\lambda_{\alpha}^{\:\beta\mu\nu}$ and $\tilde{\lambda}_{\alpha}^{\,\,\,\mu\nu}$ are two Lagrange multipliers that are useful to indicate two
additional constraints for the flatness and torsion-free conditions of spacetime. Also, note that throughout the manuscript we are working in the natural units with $c=1$ and $G=1$. In particular, by choosing $f(Q) = -Q$, one recovers the symmetric teleparallel equivalent of general relativity (STEGR).

It is also worth noting that the geometric structure considered here involves a torsion-free flat connection, which can be understood as arising from a coordinate transformation of the trivial connection, as discussed in \cite{coincident}. This connection may be acquired using the vector field $\xi^{\alpha}$, as follows,
\begin{equation}\label{2g}
\hat{\Gamma}^{\,\sigma}_{\,\,\,\alpha\beta}=\frac{\partial x^{\sigma}}{\partial\xi^{\mu}}\partial_{\alpha}\partial_{\beta}\xi^{\mu}
\end{equation}
Also note that $\xi^{\alpha} = \xi^{\alpha}(x^{\sigma})$ represents an invertible transformation. It is always possible to select this transformation such that the general affine connection vanishes, i.e., $\hat{\Gamma}^{\sigma}_{\:\alpha\beta} = 0$. This particular coordinates assumption is well identified as the coincident gauge. Hence, the components of the non-metricity tensor take the form $Q_{\sigma\alpha\beta} = \partial_{\sigma}g_{\alpha\beta}$.

Now, the energy-momentum tensor is formulated as follows,
\begin{equation}\label{2h}
\mathcal{T}_{\alpha\beta}\equiv-\frac{2}{\sqrt{-g}}\frac{\delta(\sqrt{-g}\mathcal{L}_m)} {\delta g^{\alpha\beta}}.
\end{equation}
The field equations describing the interactions in the $f(Q)$ gravity can be derived via applying the variational principle for the action \eqref{2f} and correspond to the metric as follows,
\begin{equation}\label{2i}
\frac{2}{\sqrt{-g}}\nabla_{\sigma}\left(f_{Q}\sqrt{-g}\,P^{\sigma}_{\,\,\alpha\beta}\right)+\frac{1}{2}f\,g_{\alpha\beta}+
f_{Q}\left(P_{\alpha\sigma\lambda}Q_{\beta}^{\,\,\,\sigma\lambda}-2Q_{\sigma\lambda\alpha}P^{\sigma\lambda}_{\,\,\,\,\,\,\beta}\right)=- \mathcal{T}_{\alpha\beta},
\end{equation}
where $f_Q=\frac{d f}{d Q}$. Furthermore, applying the variational principle for the action \eqref{2f} and correspond to the connection yields,
\begin{equation}\label{2j}
\nabla_{\sigma}\lambda_{\mu}^{\,\,\,\alpha\beta\sigma}+\lambda_{\mu}^{\,\,\,\alpha\beta}=\sqrt{-g}f_Q\,P_{\,\mu}^{\,\,\,\alpha\beta}+H_{\,\mu}^{\,\,\,\alpha\beta}
\end{equation}
where $H_{\,\mu}^{\,\,\,\alpha\beta}=-\frac{1}{2}\frac{\delta\mathcal{L}_m}{\delta\Gamma^{\mu}_{\,\,\,\alpha\beta}}$ denotes the hypermomentum tensor density. Taking into account the antisymmetric nature of the indices $\alpha$ and $\beta$ within the Lagrange multiplier terms, the previous expression reduces to the following form,
\begin{equation}\label{2k}
\nabla_{\alpha}\nabla_{\beta} \left(f_{Q}\sqrt{-g}\,P_{\,\mu}^{\,\,\,\alpha\beta}+H_{\,\mu}^{\,\,\,\alpha\beta}\right)=0.
\end{equation}
It is important to observe that varying the connection with respect to $\xi^{\sigma}$ effectively corresponds to implementing a diffeomorphism transformation such that $\delta_{\xi}\hat{\Gamma}^{\,\sigma}_{\,\,\,\alpha\beta}=-\mathcal{L}_{\xi}\hat{\Gamma}^{\,\sigma}_{\,\,\,\alpha\beta}=-\nabla_{\alpha}\nabla_{\beta}\xi^{\sigma}$, where the considered connection is torsion free and flat \cite{jimenez1}. When the hypermomentum is absent, the equations simplify to the following form \cite{jimenez2},
\begin{equation}\label{2l}
\nabla_{\alpha}\nabla_{\beta} \left(f_{Q}\sqrt{-g}\,P_{\,\mu}^{\,\,\,\alpha\beta}\right)=0.
\end{equation}
It is important to emphasize that the conservation law $\mathcal{D}_{\alpha}T^{\alpha}_{\,\,\,\,\beta}=0$ follows under the assumption of minimal coupling, where the matter Lagrangian depends only on the metric and not explicitly on the affine connection. In this case, the hypermomentum tensor density vanishes, and the connection field equation reduces to the equation \eqref{2l}. In the more general metric-affine framework, matter fields may couple directly to the connection, leading to non-vanishing hypermomentum. In such scenarios, the divergence of the energy-momentum tensor is sourced by hypermomentum contributions, and the standard conservation law is modified, for instance one can check the reference \cite{R291}.\\

For a matter distribution exhibiting the perfect fluid behavior, the energy-momentum tensor given as follows,
\begin{equation}\label{2m}
\mathcal{T}_{\alpha\beta}=(p+\rho)u_{\alpha}u_{\beta}+pg_{\alpha\beta},
\end{equation}
Here, $p$ and $\rho$ represent the standard pressure term and the usual matter energy density, respectively, while $u_{\alpha}$ denotes the four-velocity vector.\\
To proceed, we consider the spatially flat, isotropic and homogeneous FLRW metric, given by,
\begin{equation}\label{2n}
ds^2=-dt^2+a^2(t)(dx^2+dy^2+dz^2),
\end{equation}
In this metric, $a(t)$ denotes the scale factor. The associated non-metricity scalar is calculated to be $Q = 6H^2$. For a general functional form of $f(Q)$, the Friedmann-like equations taking into account the line element \eqref{2n} are derived as follows \cite{Lazkoz2019},
\begin{equation}\label{2o}
3H^2=\frac{1}{2f_Q} \left( -\rho+\frac{f}{2}  \right)
\end{equation}
\begin{equation}\label{2p}
\dot{H}+3H^2+ \frac{\dot{f_Q}}{f_Q}H = \frac{1}{2f_Q} \left( p+\frac{f}{2} \right)
\end{equation}
Here overdot denotes the differentiation with respect to the cosmic time $t$. Equations \eqref{2o} and \eqref{2p} can be reformulated in the following manner,
\begin{equation}\label{2q}
3H^2= \rho + \rho_{de}    
\end{equation}
\begin{equation}\label{2r}
 \dot{H}=- \frac{1}{2}\left[  \rho + \rho_{de} +p + p_{de} \right]   
\end{equation}
Here, $\rho_{de}$ and $p_{de}$ denote the energy density and pressure of the \textbf{effective} dark energy component, which originates from the non-metricity contributions. These quantities are given by,
\begin{equation}\label{2s}
\rho_{de}= \frac{1}{2} (Q-f) + Q f_Q
\end{equation}
and 
\begin{equation}\label{2t}
 p_{de}=-\rho_{de} - 2\dot{H} (1+f_Q+2Qf_{QQ}) 
\end{equation}
Furthermore, the continuity equations corresponding to both the usual matter and dark energy sector read as,
\begin{equation}\label{2u}
\dot{\rho} + 3H(\rho + p) = 0   
\end{equation}
and
\begin{equation}\label{2v}
\dot{\rho}_{de} + 3H(\rho_{de} + p_{de}) = 0 
\end{equation}

\section{Polytropic Equation of State}\label{sec3}
\justifying
The polytropic equation of state (EoS) has gained significant attention in cosmological modeling due to its ability to provide a unifying description of matter and dark energy components. Originally applied in astrophysics to model stellar structures, the polytropic EoS is expressed as,
\begin{equation}\label{3a}
 p=k \rho^\beta = k\rho^{1+\frac{1}{\alpha}}   
\end{equation}
where $p$ is the pressure, $\rho$ identifies as the energy density, $k$ denotes a constant, and $\alpha$ represents the polytropic index. The polytropic equation of state (EoS) serves as a versatile tool in cosmological modeling, particularly in the dark energy sense and modified gravity theories. By adjusting the polytropic index $\alpha$, this EoS smoothly interpolates between different fluid behaviors in the universe. Interestingly, for $\alpha = 1$, the equation becomes $p = k\rho^2$, which corresponds to the equation of state for a Bose-Einstein Condensate (BEC) dark matter model. In such a scenario, dark matter is considered to be in a condensed quantum state at low temperatures, and the pressure arises from short-range interactions between particles \cite{Boehmer2007}. The BEC EoS has been used to model both early and late-time cosmology, as it introduces repulsive pressure effects that can influence structure formation and cosmic acceleration. For negative values of $\alpha$, such as $\alpha = -1$, the EoS simplifies to a constant-pressure fluid, $p = k$, effectively mimicking the cosmological constant in $\Lambda$CDM. More importantly, when $k < 0$ and $\alpha = -\frac{1}{2}$, the polytropic EoS transforms into $p = -A/\rho$ (where $A = -k$), which is the standard form of the Chaplygin gas. This chaplygin model has been thoroughly studied in the literature of cosmology as it provides a unified description of dark energy and dark matter: it behaves like pressureless matter at early times and transitions to a dark-energy-sector at late times \cite{Kamenshchik2001, Bento2002}. Further, when $k < 0$ and $-1 < \alpha \leq -\frac{1}{2}$, the polytropic EoS transforms into $p = -A/\rho^\zeta$ (where $A = -k$ and $0 < \zeta \leq 1$), which is the standard form of the Generalized Chaplygin gas. Therefore, this range of behavior allows the polytropic EoS to unify and generalize multiple dark energy models, making it particularly attractive for use in modified gravity frameworks. Thus, the polytropic EoS serves as a promising tool in the study of cosmic acceleration within the realm of modified gravitational frameworks.\\

Note that, in the effective $w$CDM formalism, one can rewrite the equations \eqref{2q}-\eqref{2t} trivially as
\begin{align}
3H^2 &= \rho_m + \rho_{\mathrm{eff(DE)}}, \\
\dot{H} &= -\frac{1}{2} \left[ \rho_m + \rho_{\mathrm{eff(DE)}} + p_{\mathrm{eff(DE)}} \right], \\
\rho_{\mathrm{eff(DE)}} &= \frac{1}{2 f_Q} \left( -\rho + \frac{f}{2} \right) - \rho_m, \\
p_{\mathrm{eff(DE)}} &= \rho_m + \rho_{\mathrm{eff(DE)}} + 2 \frac{\dot{f}_Q}{f_Q} H - \frac{1}{f_Q} \left( p + \frac{f}{2} \right),
\end{align}
respectively. Here, $m$ denotes the effective pressureless matter for which $\rho_m \sim (1+z)^3$, while the effective dark energy encapsulates the contributions from modified gravity and polytropic equation of state of the fluid described by $p$ and $\rho$. It is important to note that our model considers only single matter component (neither ordinary matter nor radiation), governed by a polytropic equation of state, within an $f(Q)$ modified gravity framework. The rewriting of the equations \eqref{2q}-\eqref{2t} in an effective $w$CDM picture, with effective matter and effective dark energy contributions, is only a formal tool to investigate the dynamical dark energy behavior.\\

We examine a dynamically tested $f(Q)$ function exhibiting a power-law form, which has been shown to effectively capture the universe's transition from a matter-dominated stage to a de Sitter phase \cite{R341}.
\begin{equation}\label{3b}
f(Q)= \gamma \bigg(\frac{Q}{Q_0}\bigg)^n
\end{equation}
where $Q_0 = 6H_0^2$, and $\gamma$ along with $n$ serve as free parameters. Substituting equation \eqref{3b} into equation \eqref{2q}, we arrive at
\begin{equation}\label{3c}
\rho = \frac{(1-2n)}{2}\gamma \left( \frac{H}{H_0} \right)^{2n}  
\end{equation}
By evaluating equation \eqref{3c} at the current redshift $z = 0$, we obtain
\begin{equation}\label{3d}
\rho_0 = \frac{(1-2n)}{2}\gamma     
\end{equation}
and therefore, we have
\begin{equation}\label{3e}
\rho = \rho_0 \left( \frac{H}{H_0} \right)^{2n}    
\end{equation}
On accounting the Friedmann equation \eqref{2r} along with the considered $f(Q)$ function and the equation \eqref{3e}, we end up with the following non-linear first order ordinary differential equation,
\begin{equation}\label{3f}
\frac{\dot{H}}{H} + \frac{3}{2n} H = \frac{-3kH}{2n} \left[ \left(\frac{1}{2}-n\right)\gamma \right]^{\frac{1}{\alpha}} \left( \frac{H}{H_0} \right)^{\frac{2n}{\alpha}}
\end{equation}
Now, by transforming the variables in the operator via following relation $\frac{1}{H}\frac{dH}{dt}=\frac{dH}{dln(a)}$, the above equation becomes,
\begin{equation}\label{3g}
\frac{dH}{dln(a)} + \frac{3}{2n} H = \frac{-3kH}{2n} \left[ \left(\frac{1}{2}-n\right)\gamma \right]^{\frac{1}{\alpha}} \left( \frac{H}{H_0} \right)^{\frac{2n}{\alpha}}
\end{equation}
Clearly, this is the Bernoulli's equation in the independent variable $ln(a)$ representing e-folding time. Now performing the basic integration on the above differential equation and then transforming the variable via relation between redshift and scale factor as $a=\frac{1}{1+z}$, we end up with the following analytical notion of the Hubble function with redshift as independent variable, 
\begin{equation}\label{3h}
H(z)=H_0 \left[ (1+z)^{\frac{-3}{\alpha}} + k\left[\left(\frac{1}{2}-n\right)\gamma \right]^{\frac{1}{\alpha}} \lbrace{(1+z)^{\frac{-3}{\alpha}}-1 \rbrace} \right]^{\frac{-\alpha}{2n}}
\end{equation}
Thus, the obtained equation \eqref{3h} represents governing dynamics of our model involving five independent model parameters $H_0$, $\gamma$, $n$, $k$, and $\alpha$. For the fixed setting of model parameters $\gamma=-Q_0, n=1, k=0$ i.e, $f(Q)=-Q$ and $p=0$ one can recover the Einstein's general relativistic friedmann model solution for the dust case i.e. $H(z)=H_0(1+z)^\frac{3}{2}$. Also note that, that $\Lambda$CDM is recovered for specific parameter values $\gamma = -Q_0$, $n=1$, $\alpha=-1$, and $k \neq 0$, which imply that the effective fractional energy densities for the matter component
and the cosmological constant are $\Omega_{m0}=1+ k\left[\left(\frac{1}{2}-n\right)\gamma \right]^{\frac{1}{\alpha}}$ and $ \Omega_{\Lambda0}=-k\left[\left(\frac{1}{2}-n\right)\gamma \right]^{\frac{1}{\alpha}}$. Observe that in this limit, while the effective matter density evolves as $\approx (1+z)^3$ (standard matter), the effective cosmological constant arises from a combination of parameters from both the $f(Q)$ function and
the polytropic equation of state. However, in the case of our analysis, these five are free independent model parameters. Ofcourse these free parameters should not choose in an arbitrary manner. Therefore, in the upcoming section, we perform a rigorous statistical investigation to estimate the free parameter values along with their confidence limits.

\section{Bayesian Inference}\label{sec4}
\justify
To estimate the associated model parameters reliably, we work within the Bayesian statistical framework, that allows us to update prior assumptions in light of new observational data by constructing the posterior probability distribution. For sampling from this posterior, we invoke the Markov Chain Monte Carlo (MCMC) method of sampling utilizing the \texttt{emcee} Python package \cite{Mackey/2013}, which works with the help of an affine-invariant ensemble sampler. Unlike traditional Metropolis-Hastings algorithms that rely on a single chain and often struggle with tuning proposal distributions in complex or correlated parameter spaces, \texttt{emcee} uses a set of parallel walkers that adaptively explore the posterior more efficiently and are less sensitive to the initial conditions. This leads to faster convergence and more reliable sampling, especially in high-dimensional or anisotropic spaces. In our analysis, we apply this method to develop some realistic constraints on our statistical model using a different combinations of various datasets: the Cosmic Chronometer (CC) measurements, the Pantheon+SH0ES compilation of Type Ia supernovae, single compressed CMB observation, and latest Baryon Acoustic Oscillations (BAO) measurements from DESI DR2.

\subsubsection{Cosmic Chronometers}
\justify
The Cosmic Chronometer dataset provides $H(z)$ measurements derived from observations of massive galaxies, having the redshift interval $0.07 \leq z \leq 2.41$. These measurements are obtained using the differential age technique, which estimates the Hubble parameter based on the age difference between galaxies at slightly different redshifts \cite{CYU}. A comprehensive list of the available $H(z)$ data points can be found in reference \cite{RS}. The $\chi^2$ function used in our analysis is given by,
\begin{equation}\label{4a}
\chi_{CC}^{2}=\sum\limits_{k=1}^{31}
\frac{[H_{th}(z_{k},\theta)-H_{obs}(z_{k})]^{2}}{
\sigma _{H(z_{k})}^{2}}  
\end{equation}
Here, $\theta$ is a tupple with the entry components as the model parameters.

\subsubsection{SNIa datasets}
\justify
The Pantheon+SH0ES data points sample represents a comprehensive compilations of Type Ia supernovae observations to date, significantly enhancing the statistical power of cosmological analyses based on distance-redshift measurements. It extends the original Pantheon sample by incorporating improved calibrations, light-curve fits, and additional low- and high-redshift supernovae, resulting in a total of 1701 light curves from 1550 unique supernovae across the various redshift in the range $0.001 < z < 2.26$. These improvements allow for tighter constraints on cosmological parameters and a more refined probe for the cosmic expansion. Pantheon+SH0ES serves as a key dataset in testing dark energy models, modified gravity scenarios, and the standard $\Lambda$CDM framework. A detailed description of the dataset and its construction is provided in \cite{Brout}. Also note that as we employed the Pantheon+SH0ES compilation, which combines the 1701 spectroscopically confirmed SNe Ia of the Pantheon+ sample with absolute-magnitude calibration from Cepheid variable distance measurements provided by the SH0ES program. This calibration breaks the degeneracy between the Hubble constant $H_0$ and the absolute magnitude $M_B$, enabling a direct determination of $H_0$ from the SNIa Hubble diagram without requiring additional external priors. The corresponding $\chi^2$ function given as,
\begin{equation}\label{4b}
\chi^2_{SN}=\sum_{i,j=1}^{1701}\bigtriangleup\mu_{i}\left(C^{-1}_{SN}\right)_{ij}\bigtriangleup\mu_{j}
\end{equation}
Here, $C_{SN}=C_{stat}+C_{sys}$ represents the covariance matrix associated with the supernova measurements \cite{Brout}, and
\begin{align*}\label{4c}
\quad \bigtriangleup\mu_{i}=\mu^{th}(z_i,\theta)-\mu_i^{obs}
\end{align*} 
The distance modulus corresponding to the chosen $f(Q)$ model is computed using the following expression,
\begin{equation}\label{4d}
\mu(z)= 5log_{10} \left[ \frac{D_{L}(z)}{1 Mpc}  \right]+25
\end{equation}
where $D_{L}(z)$ denotes the luminosity distance, identified by the relation below,
\begin{equation}\label{4e}
D_{L}(z)= (1+z) \int_{0}^{z} \frac{ dx}{H(x,\theta)}
\end{equation}
Here, $\theta$ denotes the standard set of model parameters under consideration.

\subsubsection{CMB}

We employ the compressed CMB likelihood as the dynamical $f(Q)$ dark energy model along with the polytropic EoS primarily affect the late-time expansion history of the Universe and mainly alter the geometrical aspects of the CMB. The full CMB temperature polarization power spectrum contains several small scale non-geometric features, such as the low-$\ell$ power deficit and the lensing amplitude, that may be influenced by residual systematics and could bias constraints on dark energy models. For instance, Planck data alone have shown a mild ($\gtrsim 2\sigma$) preference for phantom like dark energy behavior, largely driven by the lack of large scale power. To minimize such potential biases and maintain a focus on background geometry, we rely on the compressed CMB likelihood \cite{PL1,PL2,PL3}.

In this approach, the CMB information is summarized into the three parameters $\{R, \ell_A, \omega_b\}$ \cite{PL4}, where
\begin{equation}
R = \sqrt{\Omega_{m0}} H_0 D_M(z_*), 
\qquad 
\ell_A = \pi \frac{D_M(z_*)}{r_s(z_*)},
\qquad
\omega_b=\Omega_bh^2
\end{equation}
with $D_M(z_*)$ denoting the comoving distance and $r_s(z_*)$ the comoving sound horizon at photon decoupling. Here $z_*$ is the redshift of recombination. These quantities efficiently encode the distance to the last scattering surface and the acoustic scale. The resulting $3 \times 3$ Gaussian likelihood, often referred to as the Wang-Wang distance prior likelihood, provides a robust and model independent way to incorporate early Universe geometric information without relying on the full CMB spectrum.

\subsubsection{BAO (DESI DR2)}
\justifying
We consider the Baryon Acoustic Oscillation (BAO) measurements from the DESI Data Release 2 (DR2) sample, which provide one of the most precise geometric probes of the late-time Universe. BAO originate from sound waves propagating in the tightly coupled photon-baryon plasma before recombination, leaving a characteristic comoving scale at the sound horizon at the drag epoch $r_d$ imprinted in the large scale clustering of matter. This scale acts as a standard ruler and allows direct measurements of cosmological distances that depend only on the background expansion history. The DESI DR2 sample covering low and intermediate redshifts from different galaxy tracers are listed in \cite{PL5}. These measurements constrain the transverse comoving distance,
\begin{equation}
D_M(z)=\int_0^z \frac{dz'}{H(z')}
\end{equation}
and the Hubble distance,  
\begin{equation}
D_H(z)=\frac{1}{H(z)},
\end{equation}
typically reported in the dimensionless forms $\lbrace \frac{D_M(z)}{r_d}, \frac{D_H(z)}{r_d}, \frac{D_V(z)}{r_d} \rbrace$ where the volume averaged distance is,
\begin{equation}
D_V(z)=\left[z\,D_H(z)\,D_M^2(z)\right]^{1/3}.
\end{equation}

The chi-square function is then defined as
\begin{equation}
\chi^2_{\rm BAO}
=
\Delta B^{\,T} C_{BAO}^{-1} \Delta B,
\end{equation}

where $\Delta B$ is defined as the difference between observed and theoretical BAO quantities, such as $D_M/r_d$, $D_H/r_d$, or $D_V/r_d$ and $C_{BAO}$ is the covariance matrix \cite{PL5}. Since BAO directly probe the late-time distance-redshift relation, DESI DR2 provides strong complementary constraints to CC, SN, and CMB data in testing the background dynamics of the polytropic $f(Q)$ cosmology.\\

The total $\chi^{2}$ function corresponding to all datasets is taken as follows,
\begin{equation}\label{4f}
\chi^{2}= \sum_i\chi^2_{i} 
\end{equation}

\section{Results}\label{sec5}
\subsection{Parameter Constraints}
We perform the minimization of the $\chi^{2}$ function to determine the median values of the model's free parameter. The $1\text{-}\sigma$ and $2\text{-}\sigma$ confidence contours for the model parameters invoking the uniform priors as $H_0 \in [50, 100]$, $n \in [0,2]$, $k \in [-2, 0]$, $\gamma \in [-10,  0]$, and $\alpha \in [-2, 0]$, based on the different combination of the datasets such as BAO+CMB, CC+SN, and joint BAO+CMB+CC+SN presented in Figures \eqref{f1}-\eqref{f1b}. The corresponding median values of these parameters, along with their 68\% confidence intervals, are summarized in Table \eqref{Table1}. Moreover, the  median values of $\Lambda$CDM model parameters invoking the uniform priors as $H_0 \in [50, 100]$ and $\Omega_{m0} \in [0,0.6]$ along with their 68\% confidence intervals, are summarized in Table \eqref{Table2}.
\begin{table}[H]
\begin{center}
\begin{tabular}{|c|c|c|c|c|c|c|c|}
\hline
Dataset&  $H_0$ & $n$ & $k$  & $\gamma$ & $\alpha$ & $\chi^2_{min}$ & $\chi^2_{red}$ \\
\hline 
BAO+CMB & $68.19^{+0.37}_{-0.34}$ & $0.99^{+0.02}_{-0.01}$ & $-1.56^{+0.16}_{-0.24} $ & $-5.87^{+0.71}_{-0.72}$ & $-1.20^{+0.11}_{-0.13}$ & 12.6 & 1.40 \\
CC+SN & $72.41^{+0.56}_{-0.60}$ & $1.17^{+0.42}_{-0.44}$ & $-1.21^{+0.27}_{-0.31} $ & $-5.92^{+0.53}_{-0.59}$ & $-1.48^{+0.24}_{-0.32}$ & 1675.10 & 0.96\\
CC+SN+BAO+CMB & $70.59^{+0.21}_{-0.23}$ & $0.97^{+0.02}_{-0.01}$  & $-0.98^{+0.11}_{-0.13} $ & $-5.98^{+0.41}_{-0.47}$ & $-1.77^{+0.15}_{-0.14}$ & 1687.70 & 0.96\\
\hline
\end{tabular}
\caption{Table shows the obtained median values of the polytropic $f(Q)$ model parameters corresponding to the various observational data samples.}\label{Table1}
\end{center}
\end{table}

\begin{table}[H]
\begin{center}
\begin{tabular}{|c|c|c|c|c|c|c|c|}
\hline
Dataset&  $H_0$ & $\Omega_{m0}$ &  $\chi^2_{min}$ & $\chi^2_{red}$ \\
\hline 
BAO+CMB & $69.45^{+0.57}_{-0.48}$ & $0.28^{+0.06}_{-0.09}$ & 15.9 & 1.32\\
CC+SN & $73.57^{+0.49}_{-0.36}$ & $0.32^{+0.02}_{-0.04}$ &  1684.90 & 0.97 \\
CC+SN+BAO+CMB & $72.97^{+0.21}_{-0.17}$ & $0.27^{+0.01}_{-0.03}$  & 1700.80 &  0.97\\
\hline
\end{tabular}
\caption{Table shows the obtained median values of the $\Lambda$CDM model parameters corresponding to the various observational data samples.}\label{Table2}
\end{center}
\end{table}

The $\chi^2_{\mathrm{red}}$ values reported in Tables \eqref{Table1} and \eqref{Table2} indicate that both the polytropic $f(Q)$ model and the $\Lambda$CDM model provide statistically acceptable fits to the observational datasets, with $\chi^2_{\mathrm{red}} \approx 1$ in most cases. For the BAO+CMB dataset, the $\Lambda$CDM model yields a slightly smaller value of $\chi^2_{\mathrm{red}}$, while for the CC+SN and the joint CC+SN+BAO+CMB datasets the polytropic $f(Q)$ model provides a marginally smaller value. Overall, the two models exhibit comparable goodness of fit to the observational data. Furthermore, the $\Lambda$CDM limit of the present model corresponds to $\alpha = -1$ and $n = 1$. From Table \eqref{Table1}, the best-fit values of these parameters are found to be close to these limits, although small deviations are present depending on the dataset combination. These deviations indicate that the observational data allow mild departures from the $\Lambda$CDM behavior within the framework of the polytropic $f(Q)$ cosmology, suggesting the possibility of a dynamical effective dark energy component.

\begin{widetext}
\begin{center}
\begin{figure}[H]
\includegraphics[scale=0.7]{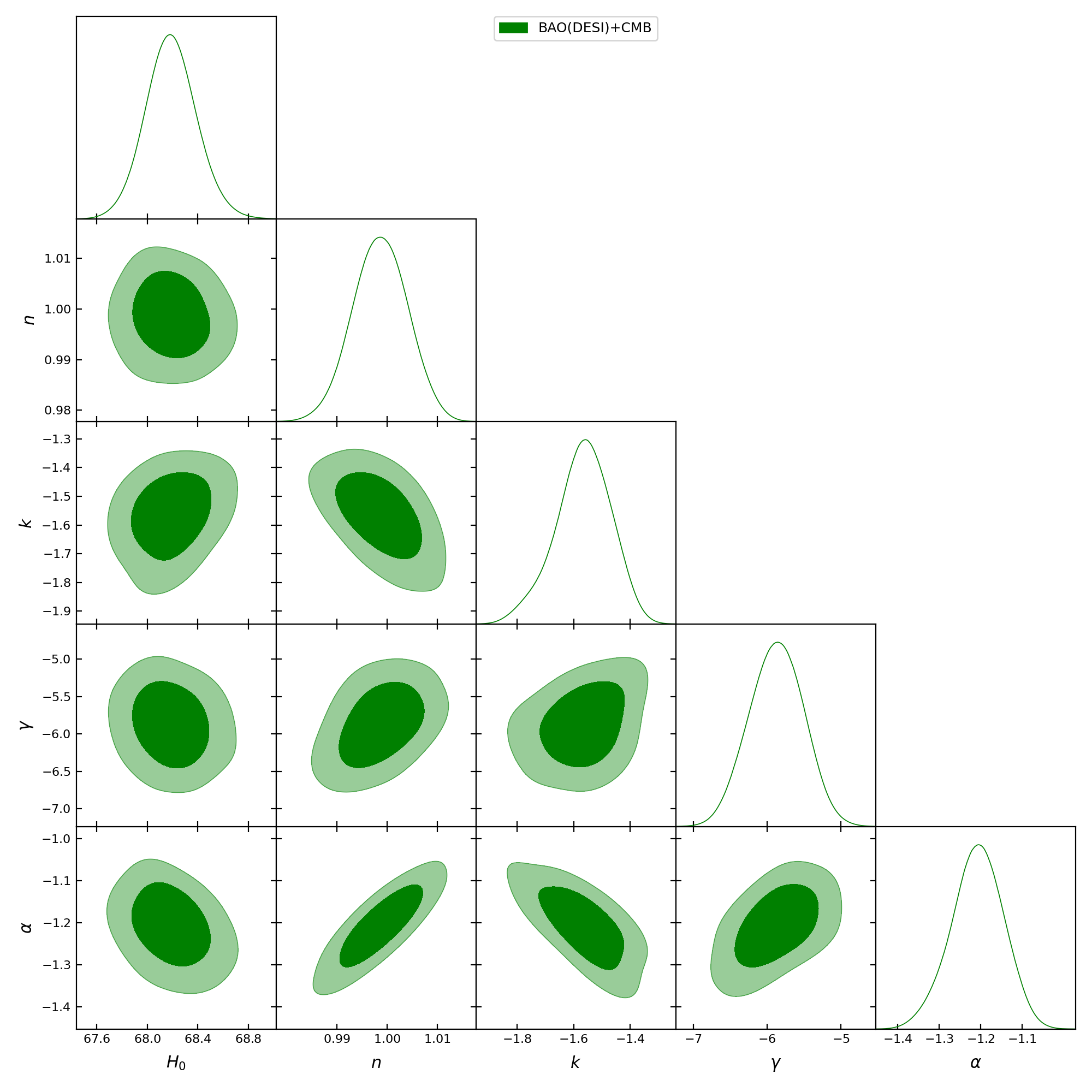}
\caption{The $1-\sigma$ to $2-\sigma$ confidence contour for the considered polytropic cosmological $f(Q)$ model using the joint BAO+CMB samples.}\label{f1}
\end{figure}
\end{center}
\end{widetext}

\begin{widetext}
\begin{center}
\begin{figure}[H]
\includegraphics[scale=0.7]{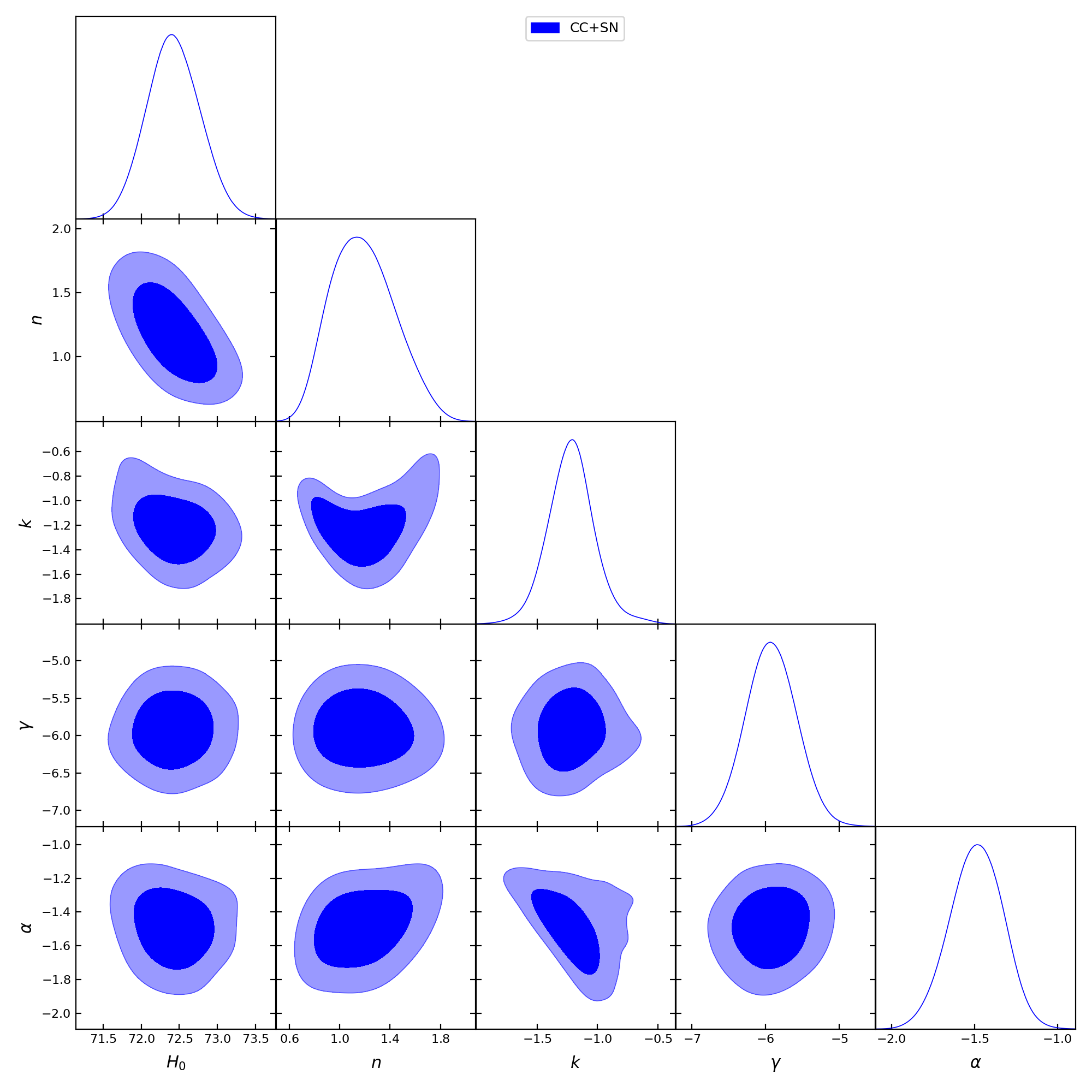}
\caption{The $1-\sigma$ to $2-\sigma$ confidence contour for the considered polytropic cosmological $f(Q)$ model using the joint CC+SN samples.}\label{f1a}
\end{figure}
\end{center}
\end{widetext}

\begin{widetext}
\begin{center}
\begin{figure}[H]
\includegraphics[scale=0.7]{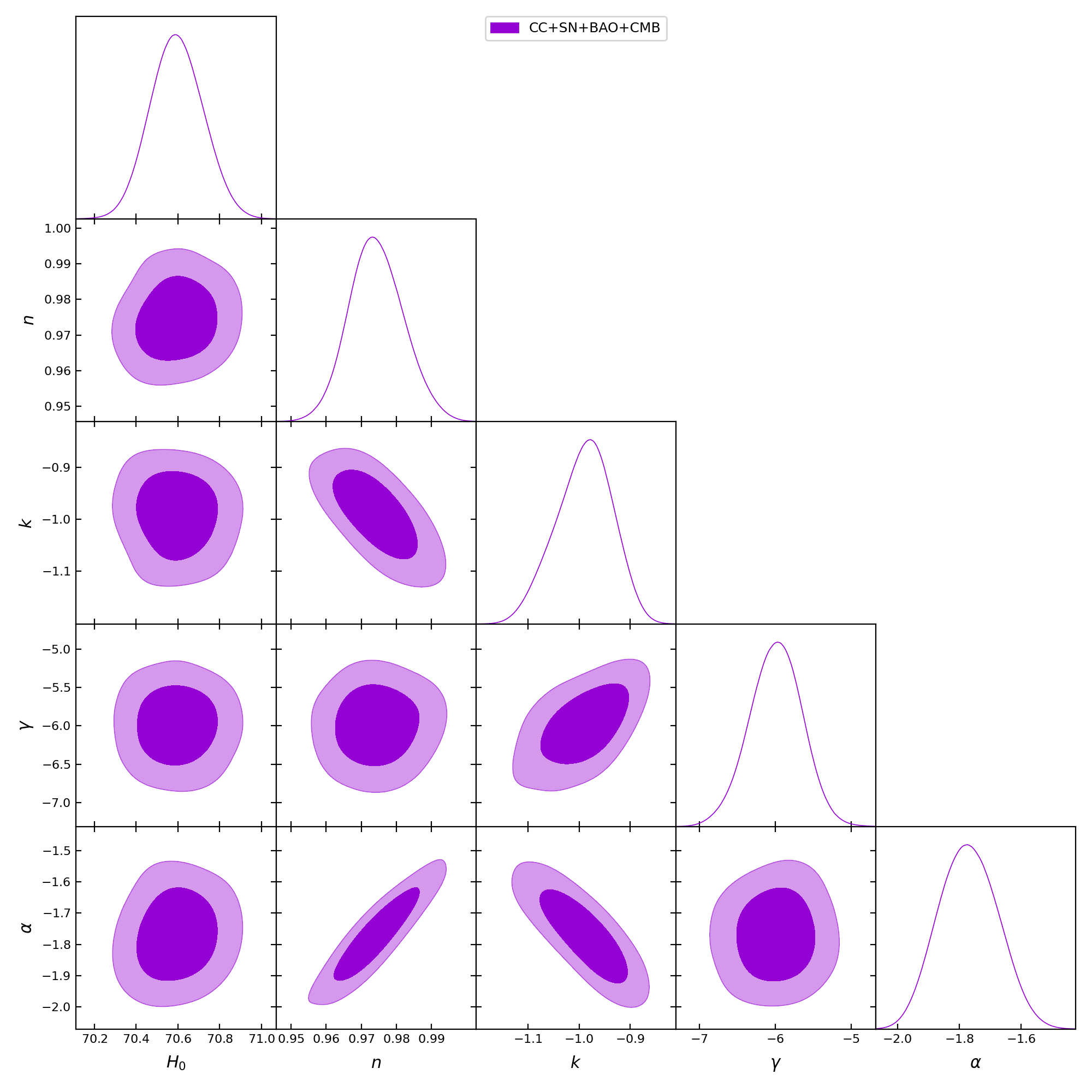}
\caption{The $1-\sigma$ to $2-\sigma$ confidence contour for the considered polytropic cosmological $f(Q)$ model using the joint BAO+CMB+CC+SN samples.}\label{f1b}
\end{figure}
\end{center}
\end{widetext}

\begin{figure}[H]
\center{\includegraphics[scale=0.51]{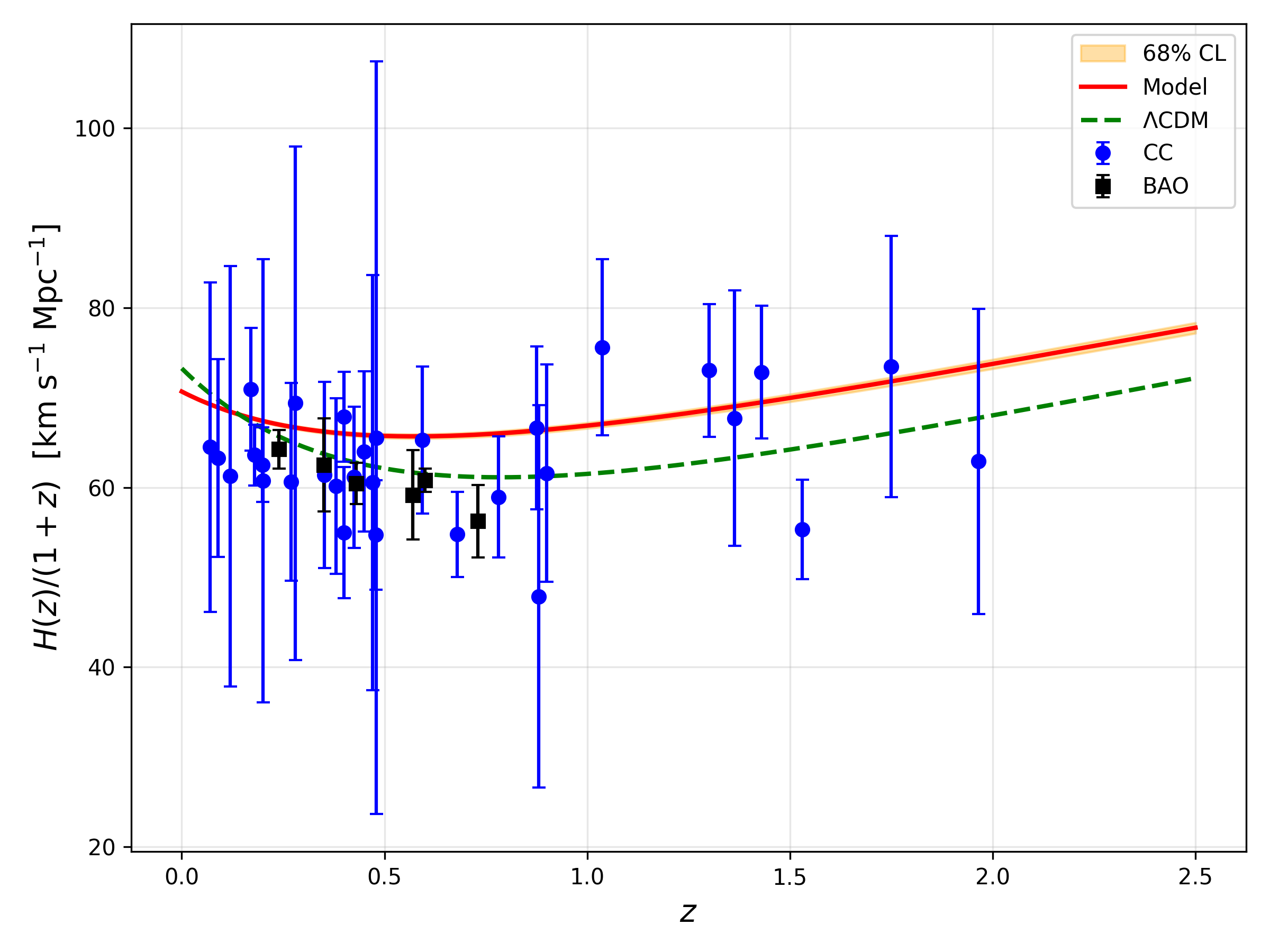}}
\caption{Plot showing the normalized expansion rate $H(z)/(1+z)$, confronted with CC and BAO observations. The red solid curve corresponds to the polytropic $f(Q)$ model while the green dashed curve corresponds to the $\Lambda$CDM model.}\label{fHz}
\end{figure}

Figure \eqref{fHz} illustrates the reconstructed evolution of the normalized expansion rate $H(z)/(1+z)$ versus redshift, confronted with Cosmic Chronometer (CC) and BAO observations. The red solid curve corresponds to the median value of the polytropic $f(Q)$ cosmological model parameters, while the shaded band represents the $68\%$ confidence interval obtained from the Bayesian analysis reported in the Table \eqref{Table1}. The dashed green curve shows the standard $\Lambda$CDM prediction for comparison using the parameter values reported in the Table \eqref{Table2}. While the CC data exhibit substantial scatter due to relatively large observational uncertainties, especially at intermediate and high redshifts, the BAO measurements provide tighter constraints at low redshift. The reconstructed expansion history remains compatible with both datasets over the entire redshift range and displays a smooth, mildly increasing behavior toward the present epoch. This trend indicates a deviation from the $\Lambda$CDM expansion rate and highlights the role of non-metricity effects and the polytropic equation of state in modifying the cosmic expansion history without spoiling agreement with observational data.

\subsection{Posterior Covariance Matrix}
\justifying

\begin{widetext}
\begin{figure}[H]
\includegraphics[scale=0.4]{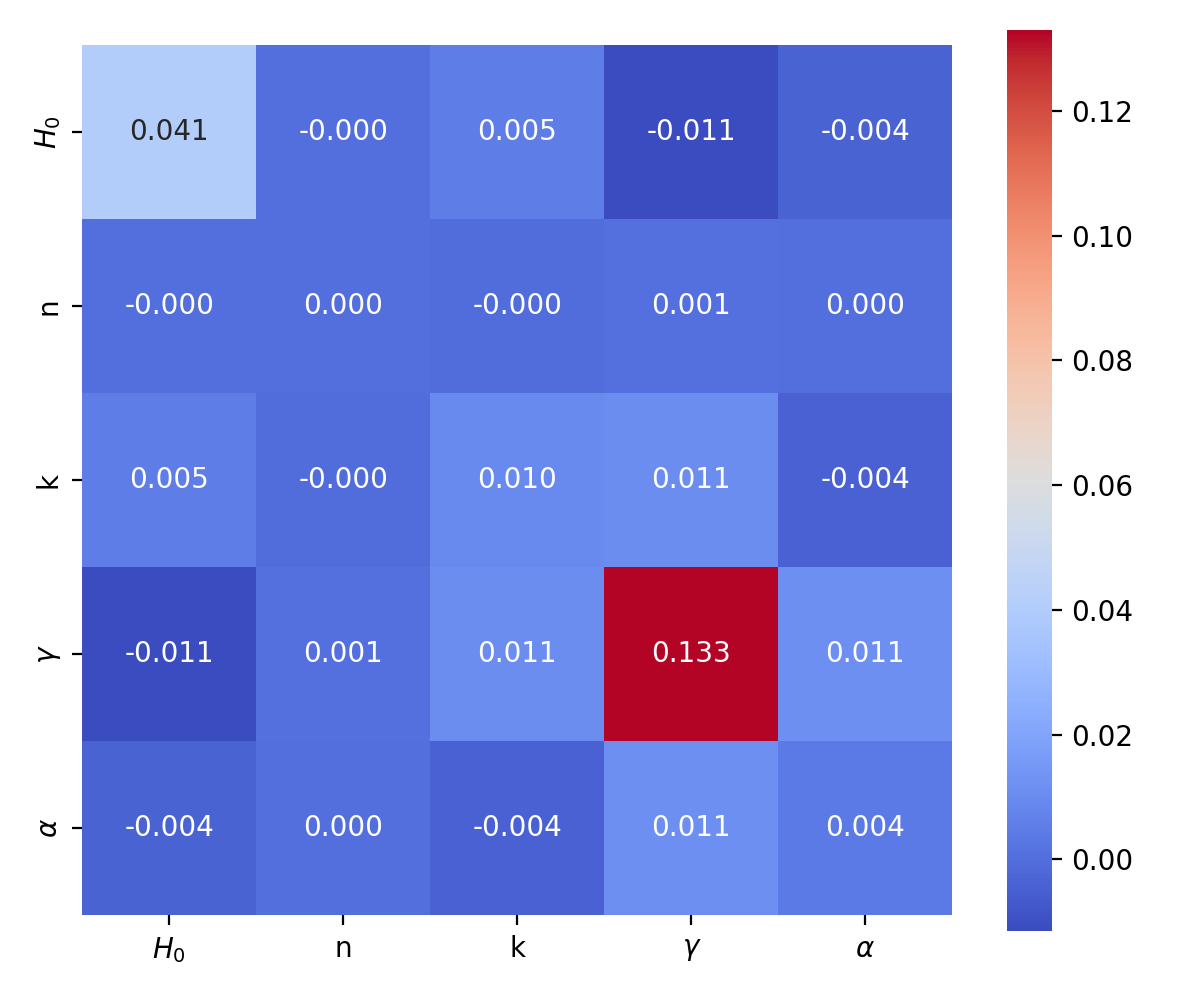}
\includegraphics[scale=0.4]{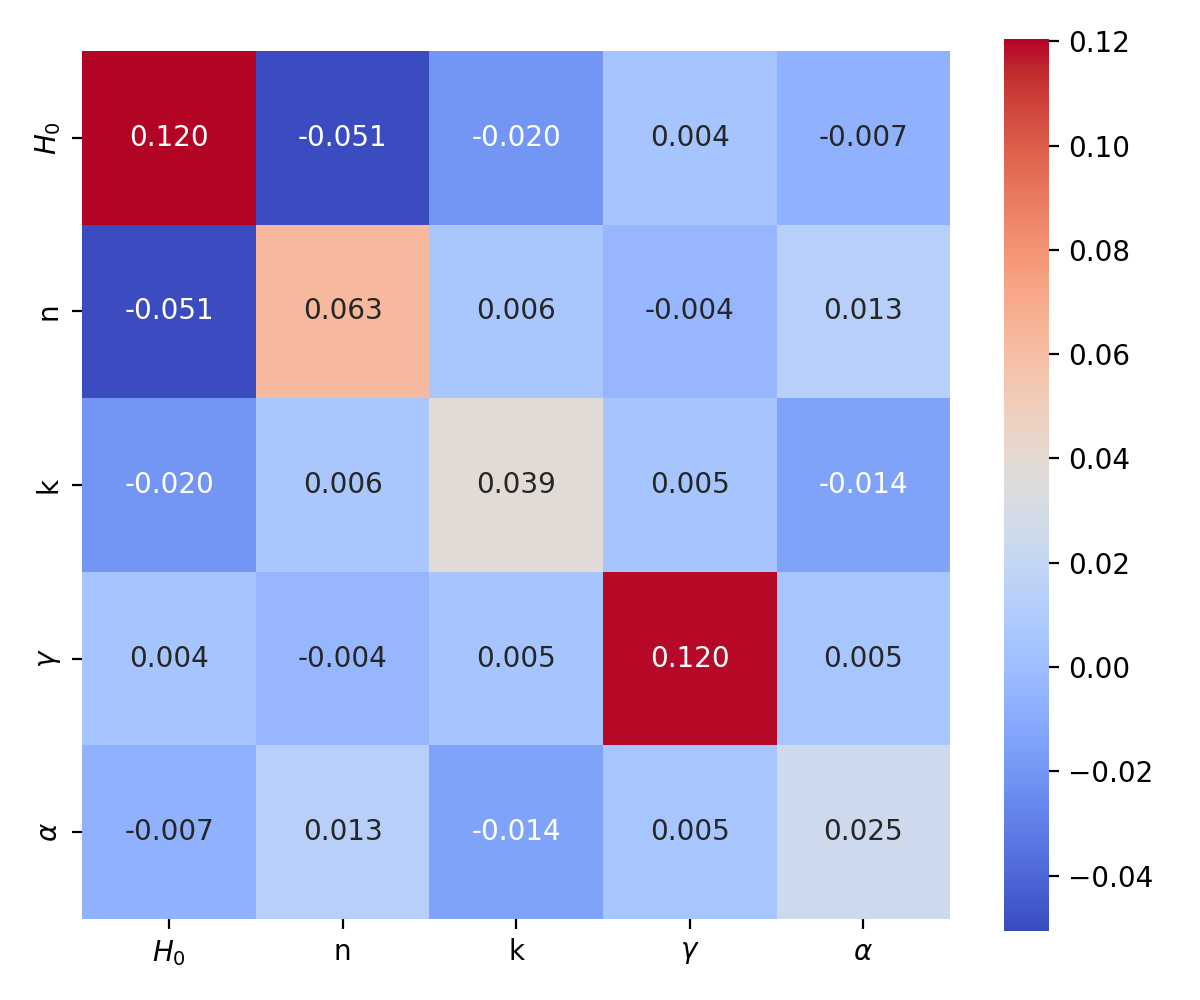}
\includegraphics[scale=0.4]{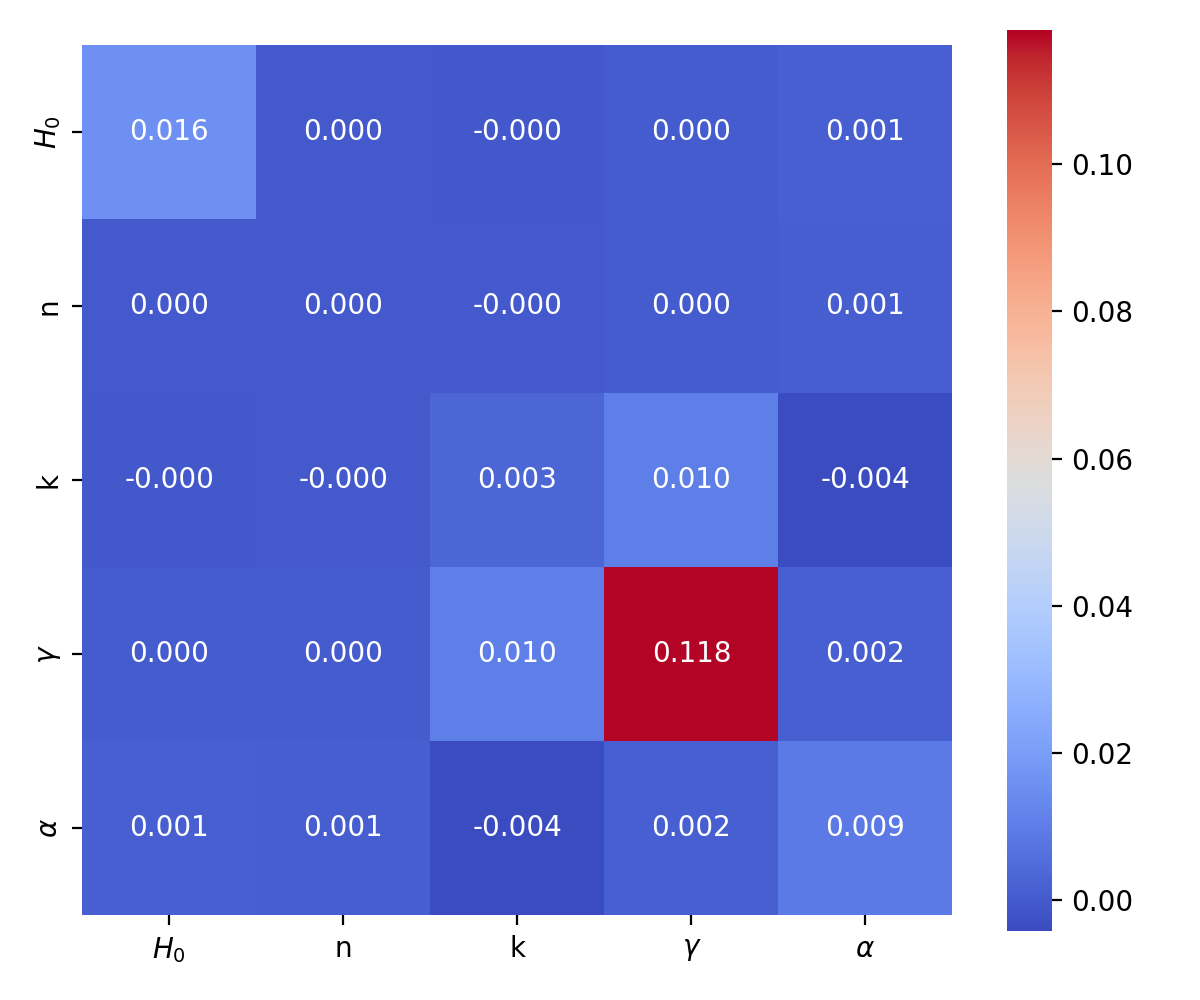}
\caption{Covariance matrices representing the relationship among the constrained model parameters corresponding to the dataset combination BAO+CMB (left), CC+SN (middle), and BAO+CMB+CC+SN (right) samples.}\label{f2}
\end{figure}
\end{widetext}

Following the MCMC simulation performed using the \texttt{emcee} sampler, we construct the posterior covariance matrices to examine the statistical relationships among the estimated model parameters. The covariance matrix retain the physical scales of the parameters and quantify how uncertainties in one parameter propagate into others, with diagonal elements representing parameter variances and off-diagonal terms indicating joint variations. These matrices are computed directly from the converged MCMC posterior samples and therefore capture the full parameter uncertainties and degeneracies allowed by the data. The covariance matrices corresponding to the three dataset combinations - BAO+CMB, CC+SN, and the joint BAO+CMB+CC+SN sample are shown in Figure \eqref{f2}. Smaller off-diagonal values indicate weaker parameter degeneracies and more independent constraints, whereas larger off-diagonal terms signal stronger coupling between parameters; positive values imply parameters tend to increase together, while negative values indicate anti-correlated behavior. The matrices show overall diagonal dominance with moderate but non-zero off-diagonal components, revealing that while individual probes constrain parameters at different levels, their combination enhances parameter interdependence in a controlled manner and significantly tightens the overall parameter uncertainties, leading to more stable and robust cosmological inference within the polytropic $f(Q)$ framework.

\subsection{Expansion Behavior}
\justifying
The deceleration parameter, denoted by $q$, is a crucial diagnostic tool in cosmology for characterizing the expansion behavior of the cosmos. It is mathematically expressed as $q = -1 - \frac{\dot{H}}{H^2}$, where $H$ is the Hubble parameter and $\dot{H}$ represents its time derivative. This dimensionless quantity essentially quantifies the rate which measures the cosmic expansion whether it is speeding up or slowing down. A positive value of $q$ corresponds to a decelerating universe, usually associated with a radiation or matter-dominated phase, while a negative value indicates accelerated expansion, which is suggested by the current observations. The transition of $q$ from positive to negative thus signifies a pivotal shift in cosmic dynamics, highlighting the onset of dark energy dominance. By analyzing the deceleration parameter dynamics across redshifts, one can effectively trace the universe's transition from deceleration in the past to acceleration in the present, offering key understanding of the evolutionary stage of the cosmic expansion and the corresponding gravitational framework.

\begin{figure}[H]
\includegraphics[scale=0.54]{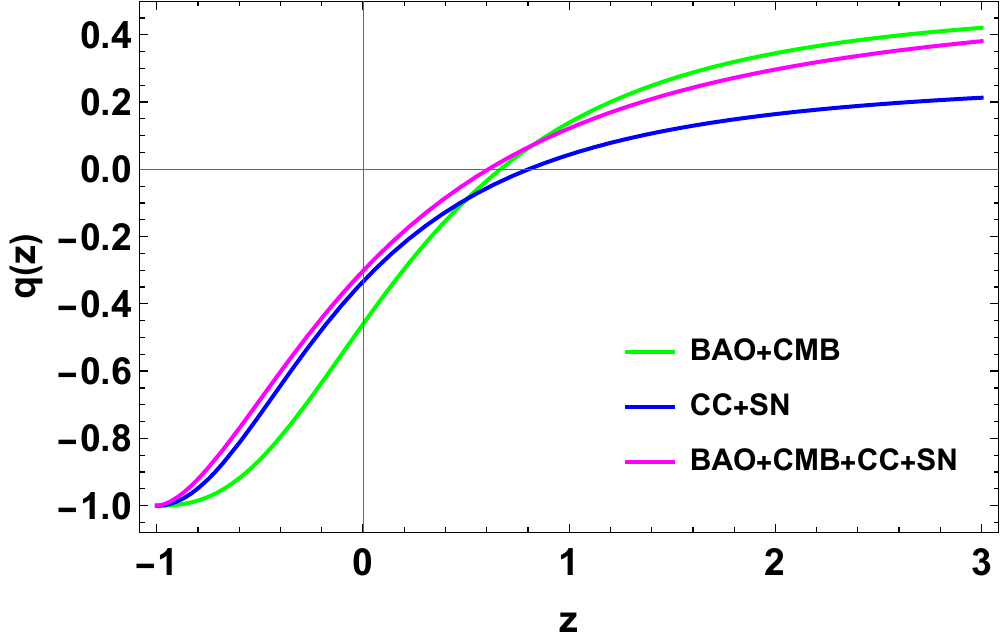}
\includegraphics[scale=0.54]{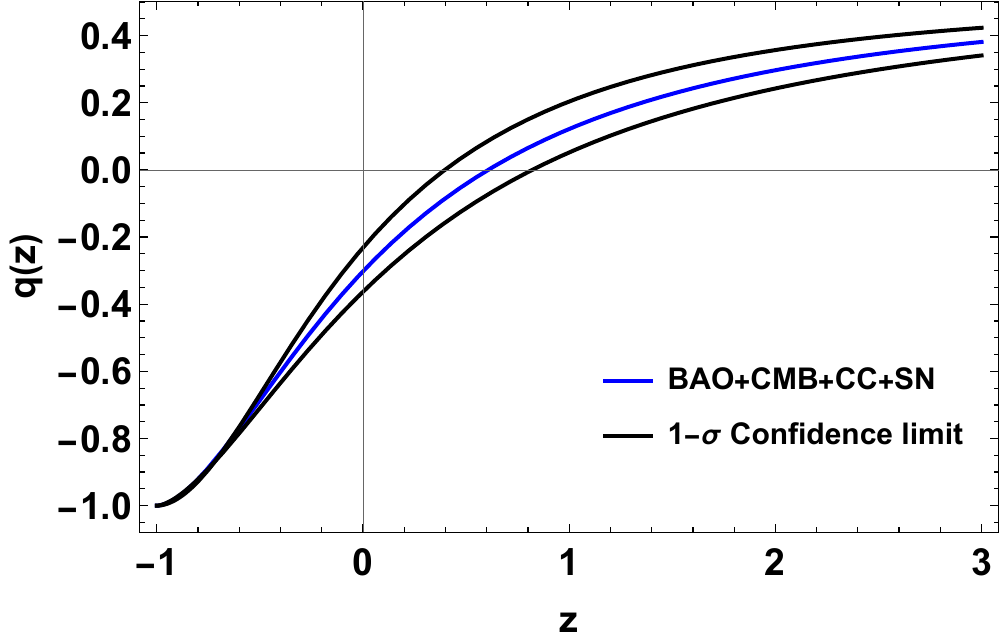}
\caption{Plot shows the comparison of the deceleration parameter behavior corresponding to the median values from the posterior distributions using three data combinations, BAO+CMB, CC+SN, and BAO+CMB+CC+SN samples (left), and for combined data along with 68\% confidence limit (right).}\label{f5}
\end{figure}
The redshift varying trajectory of the deceleration parameter $q(z)$, as obtained in Fig \eqref{f5}, exhibits a characteristic cosmological transition. For higher redshift values ($z > 0$), the trajectory emerges around $q \simeq 0.5$, consistent with the matter-dominated decelerating phase of the universe. As cosmic time progresses toward the present epoch, $q(z)$ undergoes a smooth transition from positive to negative values, indicating the onset of accelerated expansion in the recent past \cite{Betoule2014, Planck2020}. The value of the deceleration parameter at the present redshift are obtained as $q_0 \approx -0.46$ (BAO+CMB), $q_0 \approx -0.34$ (CC+SN), $q_0 \approx -0.31$ (BAO+CMB+CC+SN), with the transition redshifts $z_t \approx 0.66$ (BAO+CMB), $z_t \approx 0.80$ (CC+SN), $z_t \approx 0.60$ (BAO+CMB+CC+SN). In the future epoch ($z \to -1$), the deceleration parameter approaches to $q = -1$ asymptotically that corresponds a de Sitter-like expansion driven by an effective cosmological constant. This behaviour is well fitted with the recent observational studies and supports the scenario of a late-time acceleration phase following a prolonged decelerated epoch.
\begin{figure}[H]
\includegraphics[scale=0.54]{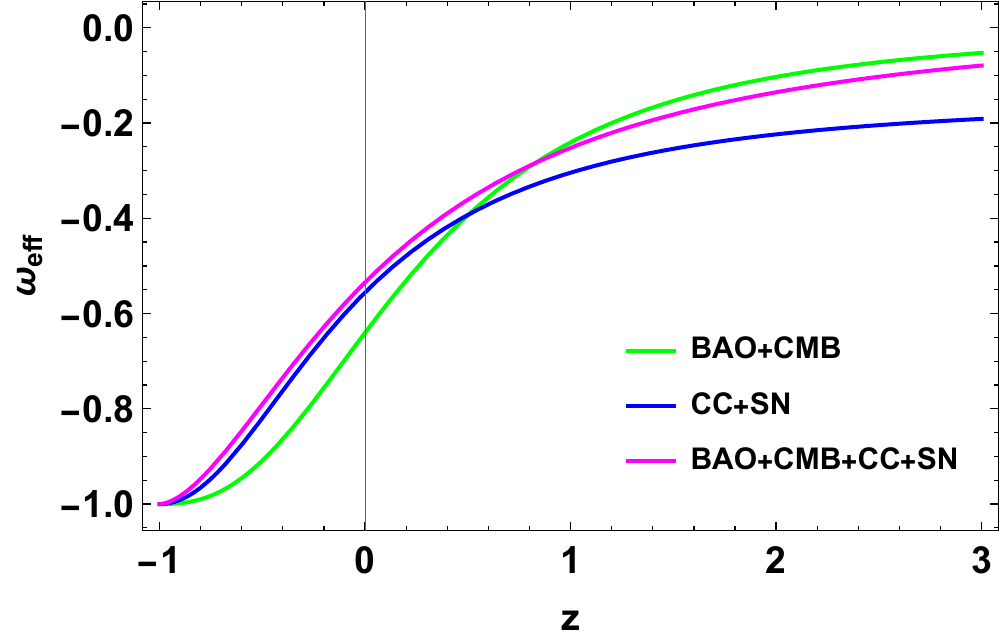}
\includegraphics[scale=0.54]{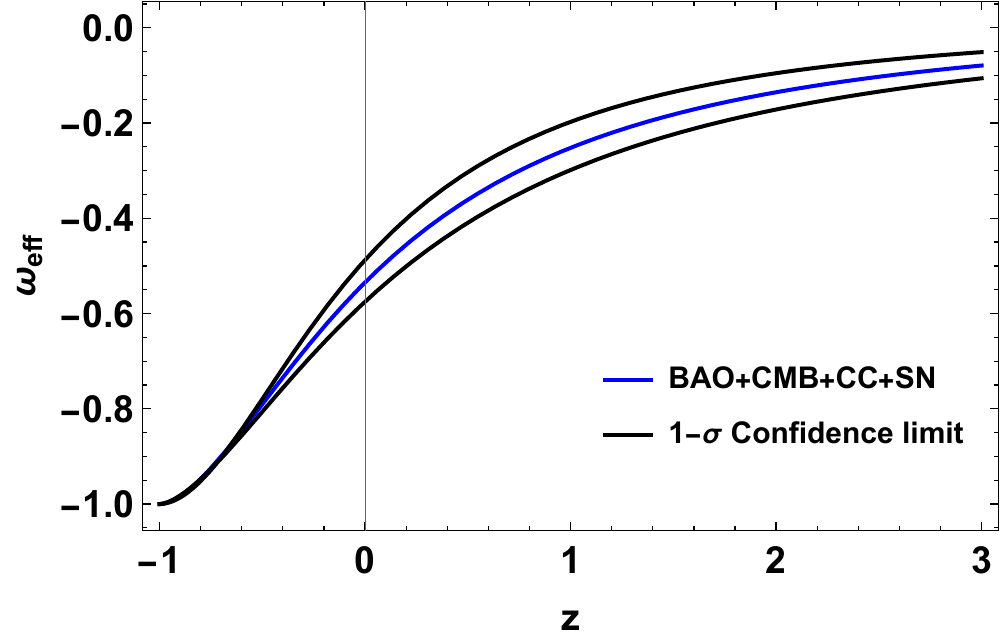}
\caption{The effective equation of state parameter behavior corresponding to the median values from the posterior distributions using three data combinations, BAO+CMB, CC+SN, and BAO+CMB+CC+SN samples (left), and for combined data along with 68\% confidence limit (right).}\label{f5a}
\end{figure}

\begin{figure}[H]
\includegraphics[scale=0.54]{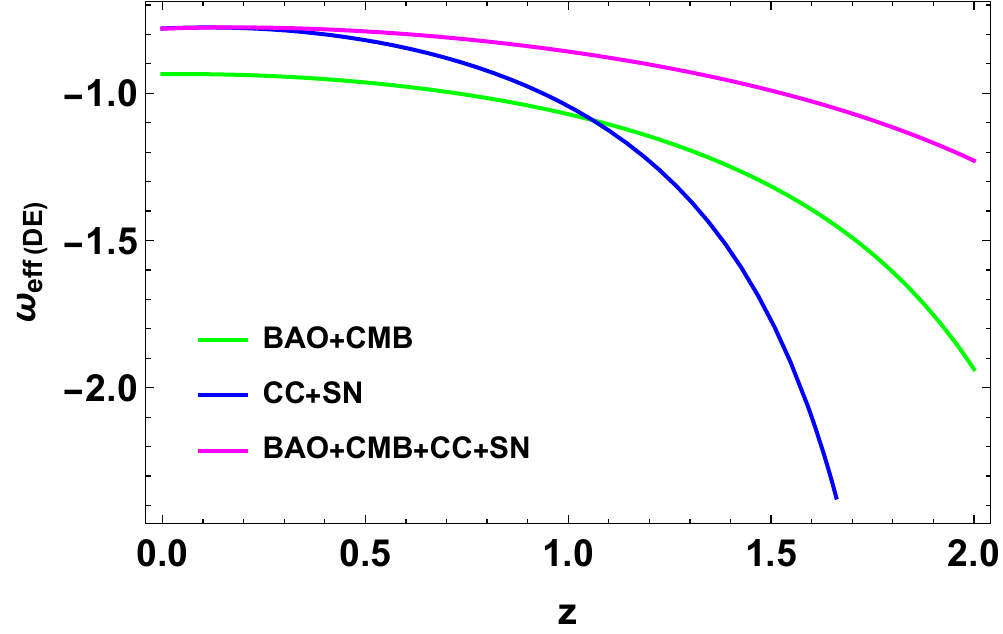}
\includegraphics[scale=0.54]{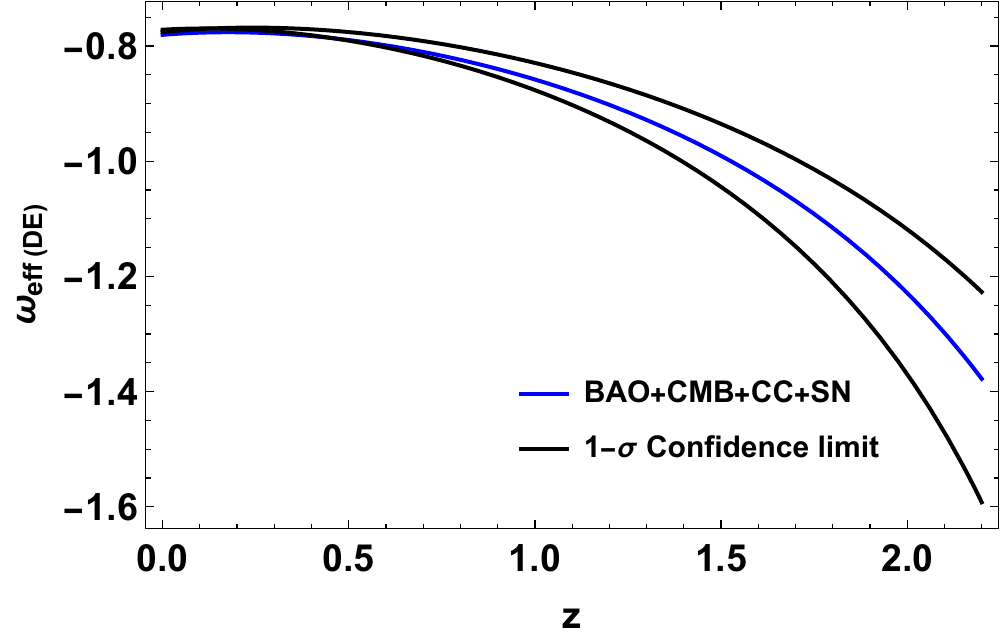}
\caption{The effective dark energy equation of state parameter behavior corresponding to the median values from the posterior distributions using three data combinations, BAO+CMB, CC+SN, and BAO+CMB+CC+SN samples (left), and for combined data along with 68\% confidence limit (right).}\label{fde}
\end{figure}

The effective equation of state parameter (EoS), denoted by $\omega_{eff}$, plays a vital role in cosmology to describe the evolutionary phases of the Universe, whereas the effective dark energy equation of state parameter is crucial for characterizing the underlying nature of dark energy. These parameters mathematically defined as $\omega_{eff} = -1 - \frac{2\dot{H}}{3H^2}$ and $\omega_{eff(DE)} = \frac{p_{eff(DE)}}{\rho_{eff(DE)}}$, where $H$ is the Hubble parameter and $\dot{H}$ represents its time derivative. The $-1 < \omega_{eff(DE)} < -\frac{1}{3}$ corresponds to a quintessence region, $ \omega_{eff(DE)} < -1$ represents the phantom behavior, while the value $\omega_{eff(DE)} = -1$ corresponds to the $\Lambda$CDM EoS. The behavior of EoS parameters shown in Figures \eqref{f5a} and \eqref{fde} provides strong evidence in favor of late-time cosmic acceleration. By analyzing the effective EoS parameter dynamics across redshifts, one can effectively trace the universe's transition from deceleration in the past to acceleration in the present. Moreover, the behavior of the effective dark energy component towards large negative values indicates the dominance of an effective dark energy like component arising naturally within the polytropic $f(Q)$ framework. Interestingly the behavior of $\omega_{eff(DE)}$ is quite similar as obtained in the DESI DR2 Results (see figure (12) \cite{PL5}).

\subsection{Statefinder Parameters}
\justifying
The statefinder diagnostic is a powerful geometrical tool introduced to characterize and distinguish between various dark energy models beyond the standard $\Lambda$CDM framework \cite{Sahni2003, Alam2003}. Defined in terms of the scale factor $a(t)$ and its higher derivatives, the statefinder parameters $(r, s)$ are given by $r = \frac{\dddot{a}}{aH^{3}}$ and $s = \frac{r - 1}{3(q - 1/2)}$, where $H$ is the Hubble parameter and $q$ is identified as the deceleration parameter. These parameters allow for a model-independent comparison of cosmological models by mapping them into the $(r, s)$ plane, where the $\Lambda$CDM epoch is identified by a fixed point $(r, s) = (1, 0)$. Deviations from this point favor the departure of the underlying model from the cosmological constant scenario, enabling an effective discrimination between quintessence, Chaplygin gas, modified gravity scenarios, and other exotic dark energy candidates. The statefinder formalism has thus become an important complement to conventional cosmological diagnostics, offering deeper view of the dynamical behavior of the universe's accelerated expansion. The expression for the statefinder parameters in terms of the Hubble parameter and its derivatives are given as \cite{Alam2003},

\begin{equation}\label{nds1}
r(x) = 1 - 2\frac{H'}{H}x + \left\{ \frac{H''}{H} + \left(\frac{H'}{H}\right)^2 \right\} x^2
\end{equation}
and
\begin{equation}\label{nds2}
s(x) = \frac{r(x) - 1}{3\left(q(x) - \frac{1}{2}\right)}
\tag{32}
\end{equation}
where $x=1+z$ and $H'$ denotes the derivative with respect to redshift $z$.

\begin{figure}[H]
\includegraphics[scale=0.5]{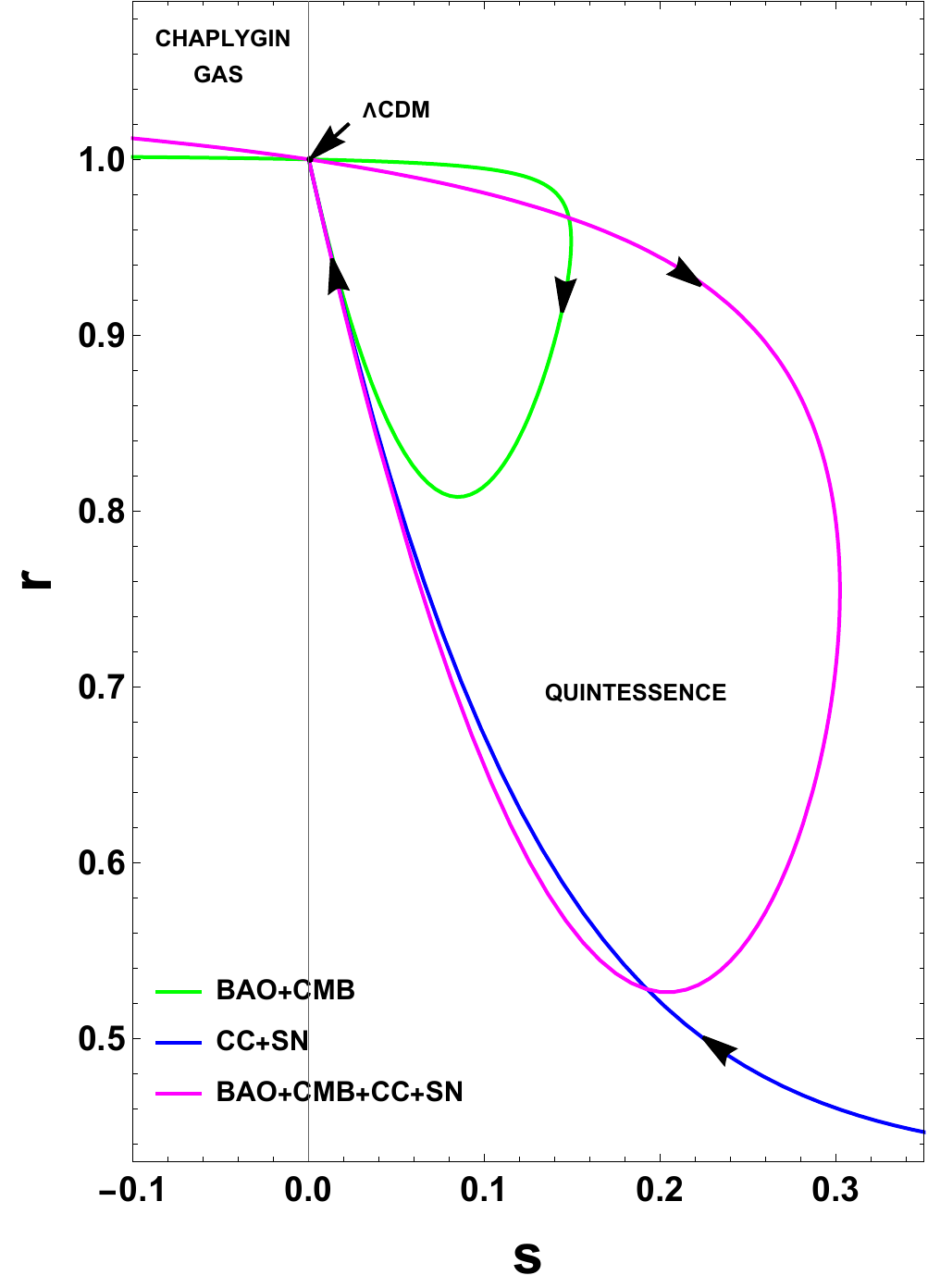}
\includegraphics[scale=0.54]{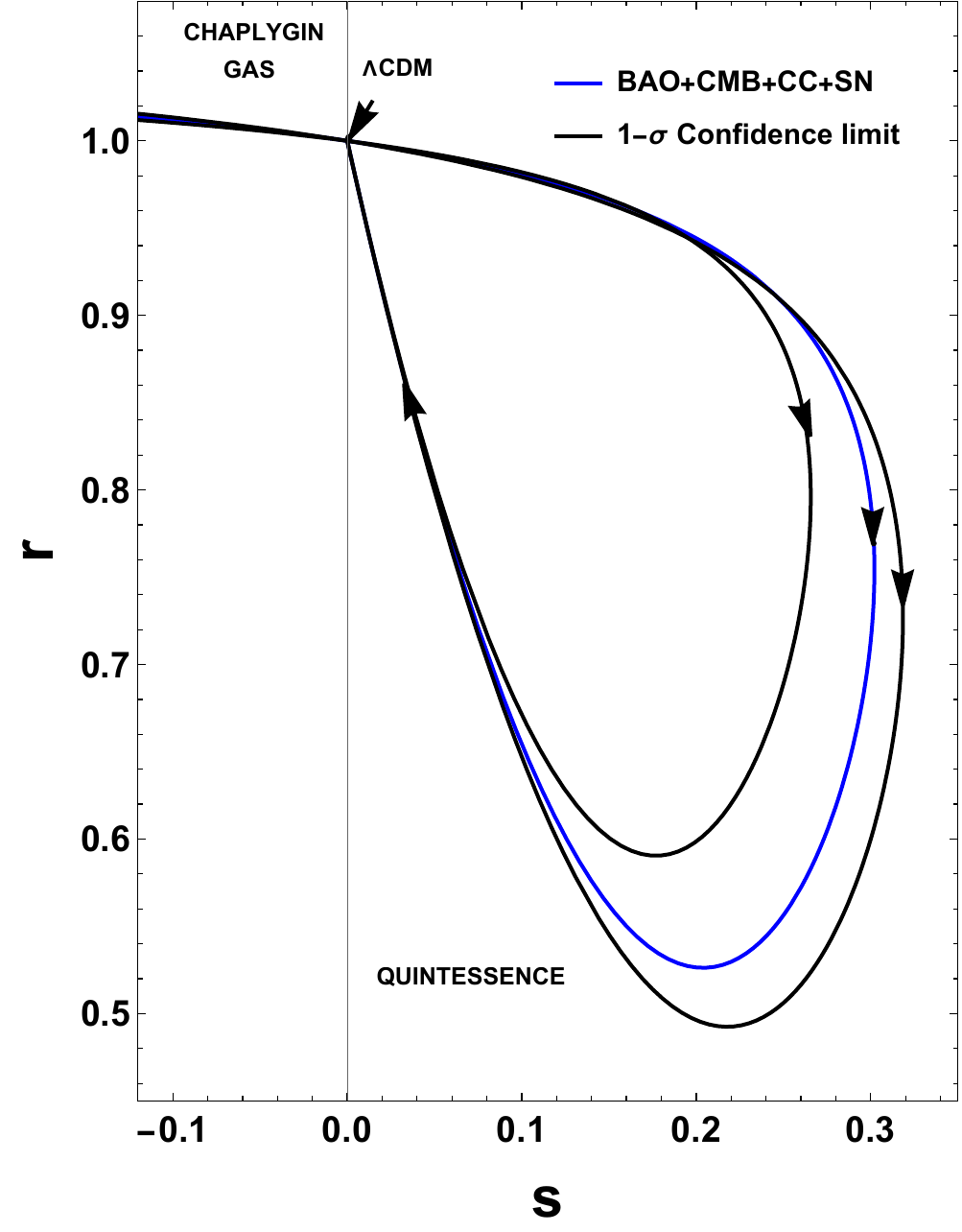}
\caption{Behavior of the statefinder parameters corresponding to the median values from the posterior distributions using three data combinations, BAO+CMB, CC+SN, and BAO+CMB+CC+SN samples (left), and for  combined data along with 68\% confidence limit (right).}\label{f6}
\end{figure}
The statefinder diagram in Fig \eqref{f6} traces the dynamical nature of the model in the $(r,s)$ plane and it obtained by evaluating $r-s$ values using the Hubble function obtained in the equation \eqref{3h}. At early times, the trajectory starts in the quadrant where $r>1$ and $s<0$, a region typically linked to a strongly decelerating universe with a dominance of non-standard matter or stiff-like behaviour. As the expansion evolves, the curve moves into the domain $r<1$ and $s>0$, signaling a shift away from the $\Lambda$CDM fixed point and the emergence of a dark energy component whose properties resemble quintessence or other evolving fields. The present value of the statefinder parameters are obtained as $(r_0,s_0)=(0.52,0.19)$ (BAO+CMB), $(r_0,s_0)=(0.49,0.19)$ (CC+SN), and $(r_0,s_0)=(0.59,0.18)$ (BAO+CMB+CC+SN). In the far future, the trajectory approaches the point $(r=1, s=0)$, corresponding to a pure cosmological constant and a de-Sitter phase of accelerated expansion. This smooth progression reflects the physical picture of the cosmos transitioning from an early decelerated phase to a dark energy driven accelerating epoch, eventually settling into a stable late-time state.

\subsection{$H_0$ Tension Analysis}
\justifying
To assess whether the proposed polytropic $f(Q)$ cosmology alleviates the Hubble constant tension, we perform a statistically consistent internal dataset comparison within each cosmological model. Rather than comparing with external determinations of $H_0$, we evaluate the tension between early and late-universe probes using identical dataset combinations. This ensures that the comparison is free from assumptions about external cosmological models and minimizes the impact of possible systematic differences between experiments. For two independent dataset combinations $A$ and $B$, the tension in the Hubble constant is quantified in units of standard deviations as,
\begin{equation}\label{T}
T = \frac{|H_0^{(A)} - H_0^{(B)}|}
{\sqrt{\sigma_A^2 + \sigma_B^2}},
\end{equation}
here $H_0^{(A)}$ and $H_0^{(B)}$ denote the median values of the Hubble constant obtained from the two dataset combinations, and $\sigma_A$ and $\sigma_B$ are their corresponding $1\sigma$ uncertainties.

We compare the constraints from
\[
H_0(\mathrm{CC+SN}) \quad \text{and} \quad H_0(\mathrm{BAO+CMB}),
\]
since the CC+SN combination primarily probes the late-time expansion history, while BAO+CMB constrains the early-universe geometry.

\subsubsection{Hubble Tension in the $\Lambda$CDM Model}

Using the $\Lambda$CDM model constraints obtained in Table \eqref{Table2}, we have
\begin{align}
H_0^{\Lambda{\rm CDM}}(\mathrm{CC+SN}) &= 73.57^{+0.49}_{-0.36}
\;\mathrm{km\,s^{-1}\,Mpc^{-1}}, \label{eq:H0_LCDM_late}\\
H_0^{\Lambda{\rm CDM}}(\mathrm{BAO+CMB}) &= 69.45^{+0.57}_{-0.48}
\;\mathrm{km\,s^{-1}\,Mpc^{-1}}. \label{eq:H0_LCDM_early}
\end{align}
The absolute difference between the two determinations is,
\begin{equation}
\Delta H_0^{\Lambda{\rm CDM}} = 73.57 - 69.45 = 4.12
\;\mathrm{km\,s^{-1}\,Mpc^{-1}}.
\end{equation}
The combined uncertainty is,
\begin{equation}
\sigma_{\rm comb} =
\sqrt{(0.36)^2 + (0.57)^2} \simeq 0.67
\;\mathrm{km\,s^{-1}\,Mpc^{-1}}.
\end{equation}
Substituting into the equation \eqref{T}, the resulting tension is
\begin{equation}
\mathcal{T}_{\Lambda{\rm CDM}}
\simeq \frac{4.12}{0.67} \approx 6.1\sigma .
\end{equation}
This confirms that within the $\Lambda$CDM framework, early and late Universe
datasets remain in strong statistical disagreement.

\subsubsection{Hubble Tension in the Polytropic $f(Q)$ Model}

We now repeat the same analysis for the polytropic $f(Q)$ cosmology using the
constraints reported in Table \eqref{Table1}. The corresponding Hubble constant estimates are,
\begin{align}
H_0^{f(Q)}(\mathrm{CC+SN}) &= 72.41^{+0.56}_{-0.60}
\;\mathrm{km\,s^{-1}\,Mpc^{-1}}, \label{eq:H0_fQ_late}\\
H_0^{f(Q)}(\mathrm{BAO+CMB}) &= 68.19^{+0.37}_{-0.34}
\;\mathrm{km\,s^{-1}\,Mpc^{-1}}. \label{eq:H0_fQ_early}
\end{align}
The difference between the late and early time determinations is,
\begin{equation}
\Delta H_0^{f(Q)} = 72.41 - 68.19 = 4.22
\;\mathrm{km\,s^{-1}\,Mpc^{-1}}.
\end{equation}
The combined uncertainty is,
\begin{equation}
\sigma_{\rm comb} =
\sqrt{(0.37)^2 + (0.60)^2} \simeq 0.71
\;\mathrm{km\,s^{-1}\,Mpc^{-1}}.
\end{equation}
Hence, the corresponding tension level is,
\begin{equation}
\mathcal{T}_{f(Q)}
\simeq \frac{4.22}{0.71} \approx 5.9\sigma .
\end{equation}

\subsubsection{Physical Interpretation}

The above results indicate that, although the polytropic $f(Q)$ model provides
an excellent fit to late-time cosmological data and allows for a dynamically effective dark energy sector, it does not fully reconcile the discrepancy between early and late Universe determinations of the Hubble constant. In fact, the level of internal tension between the CC$+$SN and BAO$+$CMB datasets indicates a mild relieve but comparable to that found in the $\Lambda$CDM model.

Nevertheless, the physical origin of the tension differs in the two frameworks. In the polytropic $f(Q)$ model, the modified non-metricity sector and the polytropic equation of state alter the late-time expansion history without
significantly affecting the early Universe geometric constraints. This suggests that an additional mechanisms such as scale-dependent gravitational effects, interactions in the dark sector, or extensions beyond background level dynamics are required to achieve a full resolution of the Hubble tension.
\begin{figure}[H]
\center{\includegraphics[scale=0.51]{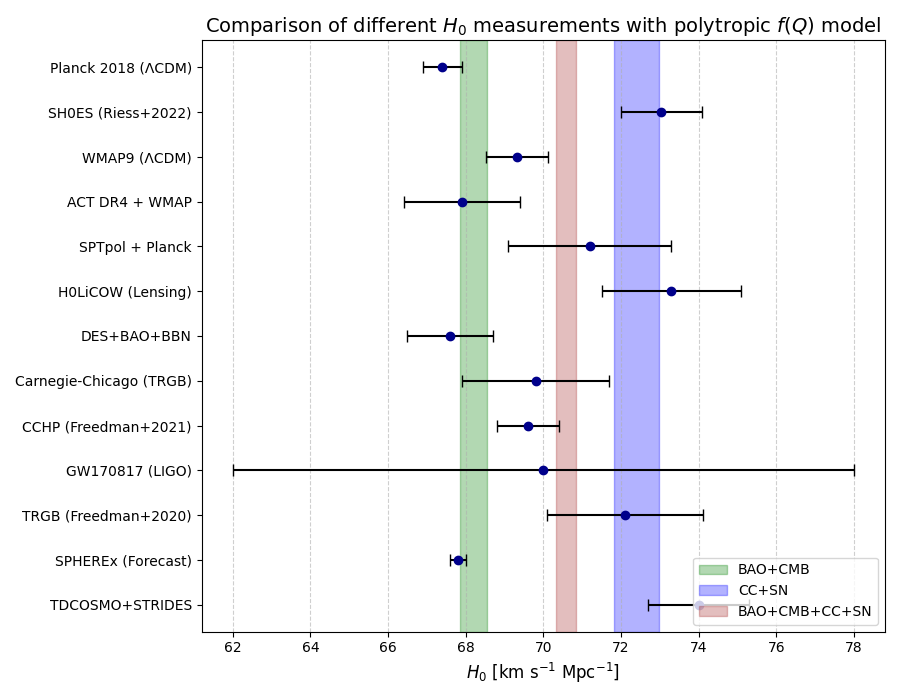}}
\caption{Plot showing the comparison among the different $H_0$ values}\label{f7}
\end{figure}
Also, we have presented a comparison between various independent determinations of the Hubble constant and the constraints obtained within the polytropic $f(Q)$ cosmological model summarized in Figure \eqref{f7}. Early-time probes such as BAO+CMB consistently favor lower values of $H_0$, while late-time measurements dominated by CC+SN prefer higher estimates, reflecting the Hubble tension. Importantly, when all datasets are combined, the inferred value of $H_0$ shifts toward an intermediate regime, lying between the early and late Universe determinations. This result indicates that the polytropic $f(Q)$ framework naturally drives joint constraints toward statistically balanced values, softening the discrepancy without significantly altering early Universe physics, although a complete resolution of the tension is not achieved.\\

\section{Concluding Remarks}\label{sec6}
\justifying
In this article we attempt to investigate the nature of the missing fluid of the universe. In order to do so, we have considered a polytropic equation of state that is more generic than the traditional equation of state, also, it is found to be very useful in the study of astrophysical objects. Moreover, particular parameter limits of the polytropic EoS naturally reduce to well-known cosmological models: For negative index ($\alpha = -1$), it reproduces a constant-pressure fluid equivalent to $\Lambda$CDM. For $\alpha = -\tfrac{1}{2}$, it yields the Chaplygin gas EoS \cite{Kamenshchik2001}, which unifies dark matter and dark energy.  For $-1 < \alpha \leq -\tfrac{1}{2}$, it leads to the generalized Chaplygin gas \cite{Bento2002}, another well-studied dark energy model. Thus, the polytropic EoS is not only mathematically flexible but also physically motivated for cosmology because it encapsulates multiple dark energy models in a single generalized framework. Thus it is quite interesting to produce some realistic constraint on the free parameter rather assuming directly a particular form of the fluid. In addition, for the background geometry we have considered a power-law $f(Q)$ cosmology which is recently proposed and found to attractive in the study of late-time cosmology.

In section \ref{sec1}, we thoroughly discussed the core motivation and phenomenology of the work, such as the $H_0$ tension, the geometric trinity, the non-metricity based $f(Q)$ cosmology, and the polytropic EoS. In section \ref{sec2}, we presented the mathematical foundation of the $f(Q)$ gravity along with the field equation and continuity equation within the flat FLRW background metric. In section \ref{sec3}, we have calculated exact cosmological solution for the power-law $f(Q)$ model incorporated with the polytropic EoS, presented in the equation \eqref{3h}.  In section \ref{sec4}, we have estimated the free model parameters, specifically $H_0$, $\gamma$, $n$, $k$, and $\alpha$, by constructing the posterior probability distribution. For sampling from this posterior, we applied the MCMC method of sampling utilizing the \texttt{emcee} Python package. The obtained parameter constraints for the polytropic $f(Q)$ model as well as the $\Lambda$CDM model using the same dataset combination are presented in Table \eqref{Table1} and Table \eqref{Table2}, in section \ref{sec5}. The contour plots for the polytropic $f(Q)$ model using BAO+CMB, CC+SN, and the joint BAO+CMB+CC+SN dataset presented in Figures \eqref{f1}-\eqref{f1b}. Also in the Figure \eqref{fHz}, we have presented the normalized expansion rate $H(z)/(1+z)$ versus redshift, confronted with Cosmic Chronometer (CC) and BAO observations. The reconstructed expansion history remains compatible with both datasets over the entire redshift range and displays a smooth, mildly increasing behavior toward the present epoch. In the further subsections, we have presented the behavior of different cosmological parameters such as the deceleration parameter, the effective EoS parameter, and the statefinder parameter using the median values from the posterior distributions using three data combinations, BAO+CMB, CC+SN, and BAO+CMB+CC+SN samples (see Figures \eqref{f5}-\eqref{f6}). The trajectory of the deceleration parameter emerges around $q \simeq 0.5$, consistent with the matter-dominated decelerating phase of the universe. The value of the deceleration parameter at the present redshift is obtained as $q_0 \approx -0.46$ (BAO+CMB), $q_0 \approx -0.34$ (CC+SN), $q_0 \approx -0.31$ (BAO+CMB+CC+SN), with the transition redshifts $z_t \approx 0.66$ (BAO+CMB), $z_t \approx 0.80$ (CC+SN), $z_t \approx 0.60$ (BAO+CMB+CC+SN). In the future epoch ($z \to -1$), the deceleration parameter approaches to $q = -1$ asymptotically that corresponds a de Sitter-like expansion driven by an effective cosmological constant. This behaviour is well fitted with the recent observational studies and supports the scenario of a late-time acceleration phase following a prolonged decelerated epoch. Moreover, the deepening of the effective EoS parameter towards large negative values at low redshifts indicates the dominance of an effective dark energy like component arising naturally within the polytropic $f(Q)$ framework. The statefinder parameter indicates non-standard matter or stiff-like behaviour at early times, whereas, as the expansion evolves, the curve moves into the domain $r<1$ and $s>0$, signaling a shift away from the $\Lambda$CDM fixed point and the emergence of a dark energy component whose properties resemble quintessence fields.

Finally, we analyzed the $H_0$ tension of the polytropic $f(Q)$ model and the $\Lambda$CDM model. We have compared the constraints from $H_0$(CC+SN) vs $H_0$(BAO+CMB), since the CC+SN combination primarily probes the late-time expansion history, while BAO+CMB constrains the early-universe geometry. Although, the level of internal tension between the CC+SN and BAO+CMB datasets indicates a mild relieve but comparable as that of found in the standard $\Lambda$CDM model. However, the physical origin of the tension differs in the two frameworks. This suggests that an additional mechanisms such as scale-dependent gravitational effects, interactions in the dark sector, or extensions beyond background level dynamics are required to achieve a full resolution of the Hubble tension. Also, we have presented a comparison between various independent determinations of the Hubble constant and the constraints obtained within the polytropic $f(Q)$ cosmological model summarized in Figure \eqref{f7}. Early-time probes such as BAO+CMB consistently favor lower values of $H_0$, while late-time measurements dominated by CC+SN prefer higher estimates, reflecting the Hubble tension. Importantly, when all datasets are combined, the inferred value of $H_0$ shifts toward an intermediate regime, lying between the early and late Universe determinations. This result indicates that the polytropic $f(Q)$ framework naturally drives joint constraints toward statistically balanced values, softening the discrepancy without significantly altering early Universe physics, although a complete resolution of the tension is not achieved. Also note that, as the present study focuses only on the background evolution of the Universe. Although this is an important starting point, a comprehensive evaluation of the proposed model demands an extension to the perturbative level. In particular, investigating linear cosmological perturbations and their implications for structure growth, the growth rate of matter fluctuations, and CMB anisotropies would be crucial for confronting the model with observational data and testing its compatibility with large-scale structure measurements. This represents a logical and valuable avenue for future research.

\section*{Data availability} There are no new data associated with this article.

\section*{Acknowledgments} \label{sec7}
There is no funding associated with this project.

\section*{Declaration of competing interest}
The author declare that they have no known competing financial
interests or personal relationships that could have appeared to influence
the work reported in this paper.


\end{document}